# ARB inactivation, ARGs and antibiotics degradation in hospital wastewater

*Kornelia Stefaniak[1], Monika Harnisz[1], Magdalena Męcik[1] and Ewa Korzeniewska[1,*]*

[1] Department of Water Protection Engineering and Environmental Microbiology, Faculty of Geoengineering, University of Warmia and Mazury in Olsztyn, Prawocheńskiego 1, 10-720 Olsztyn, Poland
**\*** Corresponding author: ewa.korzeniewska@uwm.edu.pl

## Abbreviations

| | |
|---|---|
| Amp | ampicillin |
| AR | antibiotic resistance |
| ARB | antibiotic-resistant bacteria |
| ARG | antibiotic resistance gene |
| ESBL | extended-spectrum beta-lactamases |
| LCA | life cycle assessment |
| LOD | limit of detection |
| MDR | multidrug-resistant |
| MDRB | multidrug-resistant bacteria |
| PF | photo-Fenton process |
| RB | resistant bacteria |
| ROS | reactive oxygen species |
| UV | ultraviolet |
| VBNC | viable but nonculturable |
| VRE | vancomycin-resistant *Enterococci* |
| WWTP | wastewater treatment plant |

**Abstract:**

Antibiotic resistance (AR) is one of the greatest public health challenges worldwide. Processes that allow the reduction of AR predictor of hospital wastewater has become crucial process that contributes to the protection of public health and the environment. The aim of this review article was to compare the effectiveness of various methods for treatment hospital wastewater in eliminating antibiotic-resistant bacteria (ARB) and degrading antibiotic resistance genes (ARGs) and antibiotics. A large number of studies dealing with wastewater treatment suggest that this topic is highly relevant and that new solutions are being developed to limit the spread of AR. Some wastewater treatment techniques have been in use for decades. Despite the negative effects of chlorine compounds, chlorination is still applied to eliminate ARB, ARGs, and drug metabolites. Ultraviolet (UV) radiation and ozonation have long been recognized for their treating properties. In the literature, advanced oxidation processes (AOPs) are increasingly often indicated as the most effective alternative to conventional treatment methods. Various methods for disinfecting hospital wastewater were reviewed and their environmental impact was analyzed in this article, and the results provide valuable insights for the further development of effective wastewater management strategies.

**Graphical abstract**

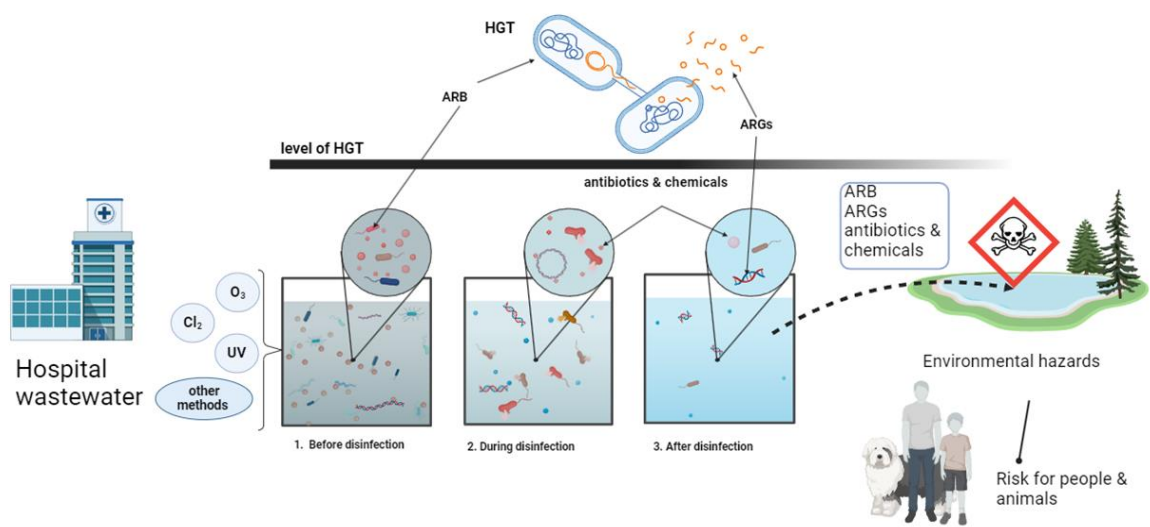

**Highlights**

[1] Methods of wastewater disinfection should base on LCA-assessed environmental impact

[2] All pre-treatment methods require individual experimental verification

[3] None method of wastewater disinfection is fully effective in removing of ARGs and ARB

# Table of contents

1.

Introduction

The discovery of antibiotics and the introduction of antimicrobials into clinical use was one of the greatest breakthroughs in human medicine (Proia et al., 2018). The data on global antibiotic consumption and the use of different classes of antibiotics clearly indicate that antimicrobial consumption has increased in all groups of countries classified by income level (ECDC, 2022). Antimicrobials are not fully metabolized by humans, and drug residues are evacuated to wastewater. These substances are present at much higher concentrations in hospital wastewater than in municipal wastewater which combines effluents from hospitals, households, industrial facilities, and storm sewers (Esther et al., 2015; Müller et al., 2018).

Hospitals generate large amounts of waste, and diverse waste streams are produced in the course of various hospital operations in diagnostic laboratories, research facilities, and patient care. Micropollutants, including antibiotics, active drug metabolites, pharmaceutical residues and microorganisms, are generally classified as hazardous waste. Pretreatment of hospital wastewater should be monitored because inadequately treated effluents can pose a threat to both public health and the environment (J. Wang et al., 2020).

Antibiotic concentrations are lower in raw wastewater reaching wastewater treatment plants (WWTPs) not only because these compounds are diluted, but also because they have been decomposed by photodegradation or hydrolysis. However, not all classes of antibiotics are susceptible to these processes (Reina et al., 2018). The antibiotics and TPs entering WWTPs can affect activated sludge (Zhang et al. 2023), potentially disrupting the biological treatment processes; in addition, pharmaceuticals could induce spreading of AR (Skandalis et al., 2021).

Due to the efficacy, availability and widespread use of antibiotics, antibiotic resistance (AR) has emerged as a global public health threat (Proia et al., 2018). Despite extensive research efforts, AR continues to pose a key challenge, and numerous studies have been undertaken to limit the spread of AR and increase public awareness about the dangers of AR for humans and animals (G. Li et al., 2020; Pokharel et al., 2019). Antibiotic resistance is directly associated with drug-resistant microorganisms that colonize living organisms and are evacuated to the environment with wastewater, where microorganisms, including environmental bacteria, can undergo various modifications. Environmental bacteria can also colonize humans through food and bathing water in lakes. Autochthonous bacterial populations may include antibiotic-resistant bacteria (ARB) that have acquired resistance mechanisms and pose a much greater threat to living organisms than environmental bacteria that are typically sensitive to most antibiotics (G. Li et al., 2020). The incidence of AR in autochthonous microorganisms in a given biotic community is directly related to the presence of ARB and antibiotic resistance genes (ARGs) in treated hospital wastewater that is discharged into surface waters without prior effective elimination. Therefore, pretreatment of hospital wastewater is crucial to limit the number of ARB and antibiotics to get into municipal wastewater. Wastewater from medical facilities is characterized by high concentrations of multidrug-resistant bacteria (MDRB) and pharmacologically active substances, including antibiotics. The specific composition of hospital wastewater significantly contributes to the promotion and spread of AR among environmental bacteria, particularly during subsequent stages of wastewater treatment. Therefore, implementing effective pretreatment technologies for hospital wastewater is essential to safeguard public health and protect the environment.

The prevalence of infections caused by clinically relevant ARB, including MDRB, continues to increase. Methicillin-resistant *Staphylococcus aureus* (MRSA), vancomycin-resistant *Enterococci* (VRE), *Enterobacteriaceae* producing extended-spectrum beta-lactamases (ESBLs), and carbapenem-resistant *Klebsiella pneumoniae* and *Acinetobacter baumanii* have been classified as priority pathogens by the World Health Organization (WHO) (Organization, 2017).

Multidrug-resistant bacteria pose one of the greatest global challenges for public health. In these microorganisms, resistance mechanisms are encoded by ARGs that are often localized on mobile genetic elements (MGEs) such as plasmids and can be transmitted to other bacteria (Delgado-Blas et al., 2022). Genetic modifications occur rapidly in microorganisms colonizing wastewater (Figure 1), particularly in hospital water which is characterized by high concentrations of antibiotics, organic matter, extracellular DNA, and ARB (Finley et al., 2013). Transmission can occur by three genetic mechanisms i.e. conjugation, transformation and transduction. Conjugation is a genetic process consisting on creating a connection by pilus between bacteria. Regardless of bacterial species, the genetic structure allows for an efficient transmission of the antibiotic resistance which is encoding on plasmids. The environment in which bacteria modification occurs intensively is rich in alive ARB and in free genetic materials that encode AR information. Transduction is subsequent process occurs by bacteriophage activity, which during infection are able to transfer DNA from previously infected bacteria by bacteriophage to another bacteria (Mosaka et al., 2023a). The wastewater environment promotes genetic mutations and adaptations to unfavorable conditions via horizontal gene transfer (HGT), which can be intensified in wastewater (Kocer et al., 2020).

A large number of substances in hospital wastewater, when discharged without pretreatment into WWTPs, contribute to higher concentrations of ARB, ARGs, and antibiotics in treated wastewater. This phenomenon is attributed to the use of ineffective treatment methods (Hubeny et al., 2021). Antibiotic resistant microorganisms can also be highly resistant to wastewater treatment in WWTPs (Hembach et al., 2017). Moreover, treated wastewater can be colonized by the most resistant strains that pose the greatest threat to the natural environment (Korzeniewska & Harnisz, 2018; Proia et al., 2018; Rodriguez-Mozaz et al., 2020).

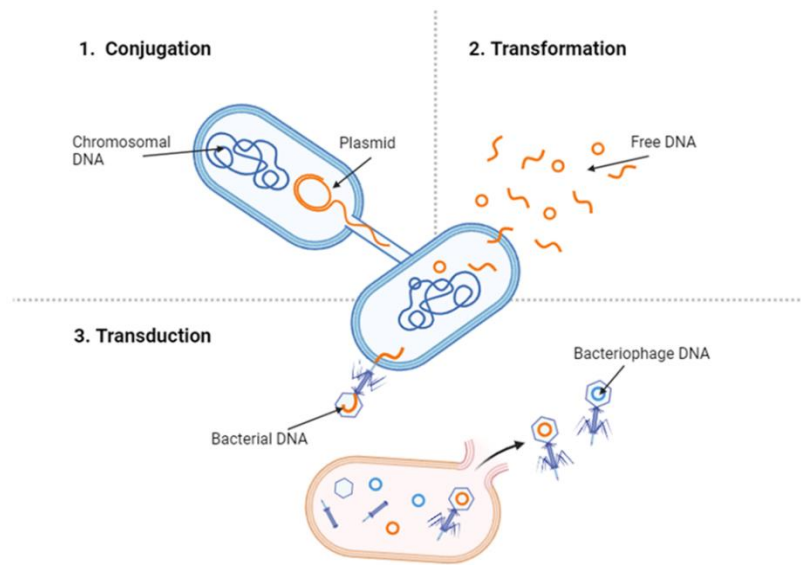

**Figure 1.** Mechanisms of antibiotic resistance gene transmission between microorganisms: 1) conjugation – exchange of genetic information via bacterial pili; 2) transformation – uptake of free environmental DNA by bacteria; 3) transduction – transfer of genetic information with the involvement of bacteriophages. Source: Mosaka et al. (2023a) with own modifications.

In 2002, the European Parliament adopted Regulation (EC) No. 1774/2002 laying down health rules concerning animal by-products not indented for human consumption. One of the main aims of this regulation was to reduce the spread of AR. Healthcare facilities that regularly come into contact with highly infectious microorganisms are legally obliged to disinfect their wastewater (The Regulation (EC) 1774/2002). According to recent research, wastewater disinfection has significantly decreased microbial counts, and these results will be discussed at length in subsequent sections of this article. Various physicochemical methods, including chlorination (Rolbiecki et al., 2023), ozonation (Cuerda-Correa et al., 2020) and ultraviolet (UV) radiation (Jäger et al., 2018), have long been used to eliminate pathogens. Ultraviolet radiation combined with other methods is one of the most popular disinfection techniques. However, side effects of hospital wastewater disinfection processes should be noted as well.

In view of the above, the aim of this article was to present the latest and the most popular disinfection methods that are applied to eliminate bacteria, ARGs (Alexander et al., 2016; Jäger et al., 2018) from hospital wastewater but also to draw attention to the beneficial side effects of disinfection which have indirect or direct impact on AR spread limitation. Research studies analyzing the concentrations of antibiotics, ARB, and ARGs in hospital wastewater and the methods for decreasing these contaminants from hospital effluents were reviewed.

This review article was motivated by the global increase of AR and the growing prevalence of infections caused by MDRB that are difficult to treat and pose a serious threat to public health. Various pretreatment techniques were compared based on their ability to reduce ARB populations and decrease concentration of ARGs and antibiotics. By compiling the knowledge about the efficacy of different pretreatment techniques, the article can guide the choice of the optimal methods that will produce the most satisfactory long-term outcomes of the protection of public health and the environment. In the present paper, various wastewater pretreatment methods were compared to achieve this goal.

## 2. Materials and Methods

A research protocol describing the research objectives, inclusion/exclusion criteria, data sources, and the browser for analyzing scientific literature was developed for the needs of this study. The literature review was based on the Preferred Reporting Items for Systematic Review and Meta-Analyses Extension for Scoping Review (PRISMA-ScR) checklist.

*2.1. Data sources*

Research articles for the review were selected according to PRISMA guidelines based on four criteria: (i) number of identified articles, (ii) number of screened articles, (iii) number of articles assessed for eligibility, and (iv) number of articles included in the final analysis. The literature search was conducted with the use of PubMed and Google Scholar databases to identify peer-reviewed articles that were published between 2002 and 2023.

*2.2. Literature search strategy*

The literature search strategy is presented in Figure S1 in the Supplementary materials. The following keywords were used in the search: "Disinfection" OR "Remove" OR "Antimicrobial resistance"; "Ozonation" OR "Chlorination" OR "UV" OR "Wastewater" OR "Hospital wastewater". The keywords were adapted to each database.

A preliminary literature search was conducted to select the keywords for the advanced search. A complementary search (including forward and backward citation search in the identified articles) was performed to determine the elements that had been omitted in the database search. The reference list was also searched manually to find research articles for the review. A total of 145 research articles and two online sources were selected. The main keywords in the reviewed scientific articles are presented in Figure S2. The reviewed articles (arranged by the year of publication) are presented in Figure S3.

The references identified during the search were imported to Mendeley (Copyright 2021 Mendeley Ltd.), and duplicates were removed. Titles and abstracts were analyzed based on adopted inclusion and exclusion criteria, and the articles selected for the review were read in their entirety.

## 3. Methods for eliminating ARB and degrading ARGs in wastewater

Wastewater disinfection processes were introduced to eliminate micropollutants, protect public health and the environment. The first reference to the effectiveness of disinfection dates back to the 19th century when chlorine was first used as a wastewater disinfectant in 1854 in London (Dymaczewski et al., 2019). Chlorination, as well as ozonation, UV radiation, and modified versions of these methods are the most popular approaches to removing microorganisms and antibiotic resistant genes from wastewater (Figure 2) (Kalli et al., 2023). More recent disinfection methods rely on peracetic acid (PAA) (Domínguez Henao et al., 2018). The selection of the optimal method is the key to successful elimination of ARB and partial elimination of ARGs. The strengths and weaknesses of each disinfection method should be analyzed to guarantee the choice of the most appropriate technique (Kalli et al., 2023; Mosaka et al., 2023a). It is important to note that the reduction in ARG concentrations is rather of a side effect of disinfection than its primary effect. This observation is supported by studies referenced later in the review.

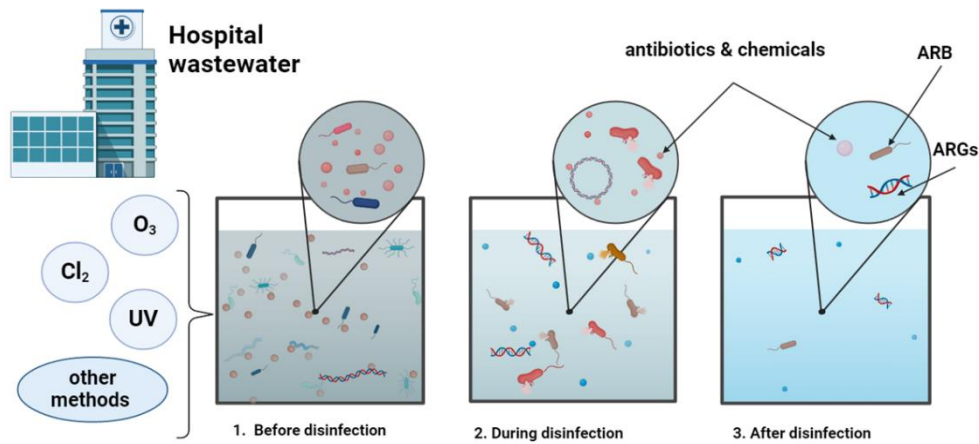

**Figure 2.** The impact of disinfectants on microorganisms colonizing hospital wastewater.(red cells – ARB sensitive to disinfection process, brown cells – ARB alive after disinfection process, blue cells – bacteria without ARGs)

*3.1. UV disinfection*

Disinfection processes that rely on UV/UV-C radiation (Δ=250-270 nm) are widely used due to their effectiveness and short contact time (Rizzo et al., 2020). UV treatments differ in their ability to eliminate ARB and degrade ARGs (Phattarapattamawong et al., 2021). UV radiation inactivates ARGs by impairing the synthesis of genetic material and inducing mutations in bacterial DNA (Barbosa et al., 2021). The mechanism of UV-radiation was illustrated on Figure 3. Numerous studies (Table 1) have demonstrated that the degradation of ARGs is determined mainly by the UV dose (Zheng et al., 2017). Zhang et al. (2015) found that a four-fold increase in the UV dose (from 62.4 mJ cm$^{-2}$ to 249.5 mJ cm$^{-2}$) increased the removal of *tetA*, *tet*B and *sul*2 gene copies from municipal wastewater by 41.7%. UV radiation was not highly effective in degrading ARGs (Y. Zhang et al., 2015), including *tet*Q and *tet*G genes (Auerbach et al., 2007), and *bla*$_{TEM}$, *qnrA*, and sul1 genes (Rafraf et al., 2016). McKinney and Pruden (McKinney & Pruden, 2012) observed that the degradation of ARGs is a more costly and energy-intensive process than ARB removal. Effective elimination of ARB required the UV dose of 20 mJ cm$^{-2}$ (to achieve a 4-5 log reduction), whereas the UV dose for ARGs inactivation was 10 to 20 times higher (to achieve a 3-4 log reduction). In addition, the cited authors found that Gram-positive VRE and MRSA were more susceptible to UV treatment than Gram-negative *E. coli* and *P. aeruginosa*. The concentrations of *tet* genes in wastewater were reduced by up to 73.5% under exposure to the UV dose of 40 mJ cm$^{-2}$, and up to 92% of ARGs were removed when the UV dose was increased two-fold. Although ARB were not effectively removed by the UV dose of 10 mJ cm$^{-2}$, an eight-fold increase in UV intensity led to their complete (100%) elimination (Zheng et al., 2017). Under exposure to the UV dose of 20 mJ cm$^{-2}$, the inactivation ratio of ARB reached 3 log for tetracycline-resistant bacteria and 4 log for heterotrophic bacteria. Tetracycline-resistant *Enterobacter* were least susceptible to UV radiation (J. J. Huang et al., 2016). In turn, Guo et al. (2013) found that the proportion of tetracycline-resistant bacteria was reduced to less than 1% after exposure to the UV dose of 5 mJ cm$^{-2}$.

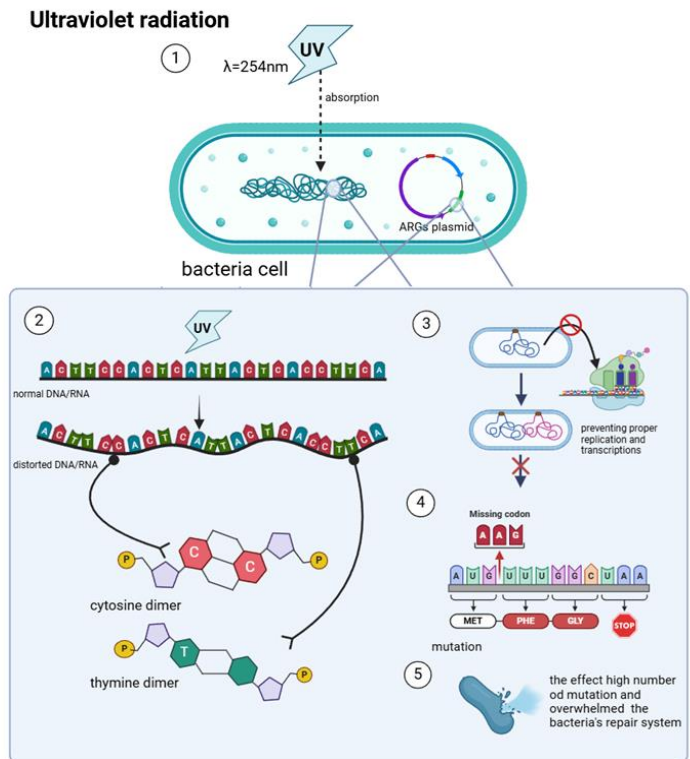

**Figure 3**. Mechanism of ultraviolet radiation in bacteria cell; 1-adsorption of UV Light *λ*=254nm (the most effective wavelengths) by bacteria cell to DNA; 2- formation thymine or cytosine dimer by formation abnormal covalent bands between adjacent pyrimidines; dimers distort the DNA structure and 3- preventing proper replication and transcription, 4- Modificated DNA cannot be properly copied what is a cause of mutation and errors during replication, 5- If the damage is extensive, the bacteria's repair system are overwhelmed. This leads to cell cycle arrest and bacteria death (created with BioRender https://www.biorender.com/ ) (Stange et al., 2019)

It could appear that all ARGs would be completely degraded when exposed to a sufficiently high UV dose, but cells are able to repair their DNA with the use of enzymes when light is not available (Oguma et al., 2013; Umar et al., 2019). *Tet* genes are more resistant to high UV doses than other ARGs (Jäger et al., 2018). It should be stressed that photoreactivation is an enzymatic process that can increase the risk of ARB reactivation (Shafaei et al., 2017). Huang et al. (2016) reported an increase in the counts of ARB and heterotrophic bacteria in samples that were exposed to the UV doses of 20 and 40 mJ cm-2 for 22 h at the temperature of 25°C. In turn, Sousa et al. (2017) found that the concentrations of ARGs (*bla*TEM and *qnr*S) in wastewater subjected to UV disinfection (254 nm, 30 min) with ozonation (30 min) reached pre-treatment levels after 3 days of storage. Wastewater is a dense matrix with complex composition, which contributes to its turbidity. UV light cannot penetrate deeper layers of turbid mixtures, which is why high reduction efficiencies are not achieved by adjusting the UV dose (Jäger et al., 2018). In a study by Zheng et al. (2017), ARB were completely inactivated by UV disinfection, whereas the concentrations of ARGs were not reduced, and the relative copy numbers of these genes even increased after the treatment. UV radiation effectively eliminated ARB and ARGs only when combined with electrocoagulation (EC) under the optimal conditions for both processes (Gomes et al., 2019).

**Table 1.** The effectiveness of modified UV radiation treatments in removing ARB and ARGs from hospital wastewater.

| UV dose | | ARGs/ARB | Reduction | References |
|---|---|---|---|---|
| **320 mJ cm$^{-2}$** | | *str*B, *tet*A, *tet*B, *aac*C2, *sul*2 | 10-20% | (H. Wang, Wang, et al., 2020) |
| **254 nm** | T= 6,9 min 5,1 mJ cm$^{-2}$ | KRE | 99% | (Azuma et al., 2024) |
| | T= 6,7 min 5,4 mJ cm$^{-2}$ | ESBL-E | | |
| | T= 1,5 min 14,9 mJ cm$^{-2}$ | MDRA | | |
| | T= 3,4 min 10,3 mJ cm$^{-2}$ | MDRP | | |
| | T= 1,9 min 17,8 mJ cm$^{-2}$ | MRSA | | |
| | T= 3 min 7,2 mJ cm$^{-2}$ | VRE | | |
| | T= 2,5 min 11,4 mJ cm$^{-2}$ | *Acinetobacter* | | |
| | T= 5,3 min 4,2 mJ cm$^{-2}$ | *Enterococcus* | | |
| | T= 7,3 min 6,0 mJ cm$^{-2}$ | *E. coli* | | |
| | T= 6,7 min 4,7 mJ cm$^{-2}$ | *Pseudomonas aeruginosa* | | |
| | T= 2,1 min 15,7 mJ cm$^{-2}$ | *Staphylococcus aureus* | | |
| **10-20 mJ cm$^{-2}$** | | MRSA, VRE, *E. coli* | <100% | (McKinney & Pruden, 2012) |
| **200-400 mJ cm$^{-2}$** | | *mec*A, *van*A, *tet*A, *amp*C | <100% | |

Despite the noticeable effect of the UV dose on the effectiveness of ARB and ARGs elimination and/or degradation, one important point should not be overlooked, namely the quality of the wastewater being disinfected. Low transparency of the wastewater (high turbidity) limits the possibility of the radiation penetrating into wastewater deeper layers, which prevents the deactivation of microorganisms (Gonz et al., 2023; Lutterbeck et al., 2020). The transparency of the wastewater is mainly influenced by the content of organic matter, which can absorb part of the radiation, weakening the effectiveness of the process. At the same time, the temperature of the wastewater is also a factor that significantly influences the effectiveness of the disinfection process. It has an indirect effect on the elimination of ARB, as the bacterial cells are more susceptible to damage at higher temperatures (Gonz et al., 2023). Nitrogen compounds present in wastewater play an equally important role. One of these is ammonium nitrogen together with its transformation products, e.g. nitrite ions ($NO_2^-$) and nitrate ions ($NO_3^-$). The ability of nitrogen compounds to absorb UV radiation limits the number of radiation waves that reach the bacterial cells and leads to the synthesis of by-products that limit the effectiveness of the process (Lutterbeck et al., 2020; Omar et al., 2024).

*3.2. Ozone disinfection*

Ozonation is a disinfection method that has long been used in wastewater treatment. Ozone $O_3$ is the highly reactive allotrope of oxygen that is produced during photodegradation of an oxygen molecule. After photodegradation, $O_3$ is synthesized from an oxygen atom and an oxygen molecule in the environment Ozone is a reactive oxygen species (ROS) that affects bacterial metabolism. Reactive oxygen species can be broadly defined as oxygen-containing radicals, including superoxide radical anions ($O_2^{•-}$), hydroperoxyl radicals ($HO_2^•$), hydroxyl radicals ($HO^•$), and alkoxy radicals ($RO^•$), as well as non-radical ROS that do not contain unpaired electrons, including ozone, singlet oxygen, and hydrogen peroxide. Reactive oxygen species oxidize the key enzymes in bacterial cells, damage bacterial DNA, inactivate and ultimately kill bacteria (Jäger et al., 2018). Being a strong disinfectant, ozone can inactivate numerous pathogens, including ARB. Gram-negative bacteria are more susceptible to ozonation because ozone increases the permeability of bacterial cell membranes (Gomes et al., 2019; Wallmann et al., 2021). Ozone leads to the oxidation of the bacterial cell wall, cell components, and genetic material (Cullen et al., 2010; Michael-Kordatou et al., 2018). The antimicrobial effectiveness of ozonation is determined by the water purity, the time of contact, and the concentration of $O_3$. The mechanism of ozonation was illustrated on figure 4. According to research (Table 2), bacteria are inactivated under exposure to 0.5 g $O_3$ $g^{-1}$ DOC (Slipko et al., 2022; von Sonntag & von Gunten, 2012), and doses higher than 0.2 g $O_3$ $g^{-1}$ DOC effectively degrade ARGs and ARB (Slipko et al., 2022). These ozone concentrations are also applied in WWTPs. Jäger et al. (2018) were used in their study 1 g ozone per 1 g DOC according to the dissolved organic carbon and a retention time of ~5min (flow rate ca. 7 $m^3$ $h^{-1}$) and they got promising results. All studied taxonomic marker genes showed reduce after ozone treatment, ranging from 98.4% for the 16S rRNA gene to below the detection limit. 99.2% and 99.7% decreases in the abundance of *E. coli* and enterococci were observed. The ozone treatment decreased 85.5 to 98.1% for all assessed antibiotic-resistance genes. Following the ozone therapy, *mec*A was undetectable. The *erm*B gene showed a reduction of 98.1%, while there was the 94.9% decrease in *sul*1, the 94.7% reduction in *int*I1, the 91% reduction in *bla*$_{TEM}$, and the 85.5% reduction in *ctx*$_{-M32}$ (Jäger et al., 2018). Water properties, in particular organic matter content, influence the effective dose of ozone which can range from 1 to 111 mg $O_3$ $L^{-1}$ (Azuma et al., 2022; Azuma & Hayashi, 2021). Ozonation is not a fully effective method of wastewater disinfection because ARGs are not completely degraded in the presence of organic substances (Lim et al., 2022). Complete elimination of ARGs and ARB requires higher ozone doses, which increases ozonation costs (Gomes et al., 2019). The ozone dose significantly affects the removal rates of ARGs and ARB. A six-fold increase in the baseline ozone dose of 13 mg $L^{-1}$ did not lead to full disinfection, but it reduced ARB counts in wastewater with high organic matter content (L. Yang et al., 2019). According to Czekalski, the effectiveness of ozonation in wastewater treatment is determined not only by the $O_3$ dose, but also by the type of targeted microorganisms (Czekalski et al., 2016). Similar observations were made by Alexander et al. (2016) who found that Gram-positive bacteria were more resistant to ozonation proccess. In turn, Slipko et al. (2022) reported that higher $O_3$ doses were required to inactivate MDRB, probably because these bacteria are more resistant to oxidative stress. However, further research is needed to confirm these observations.

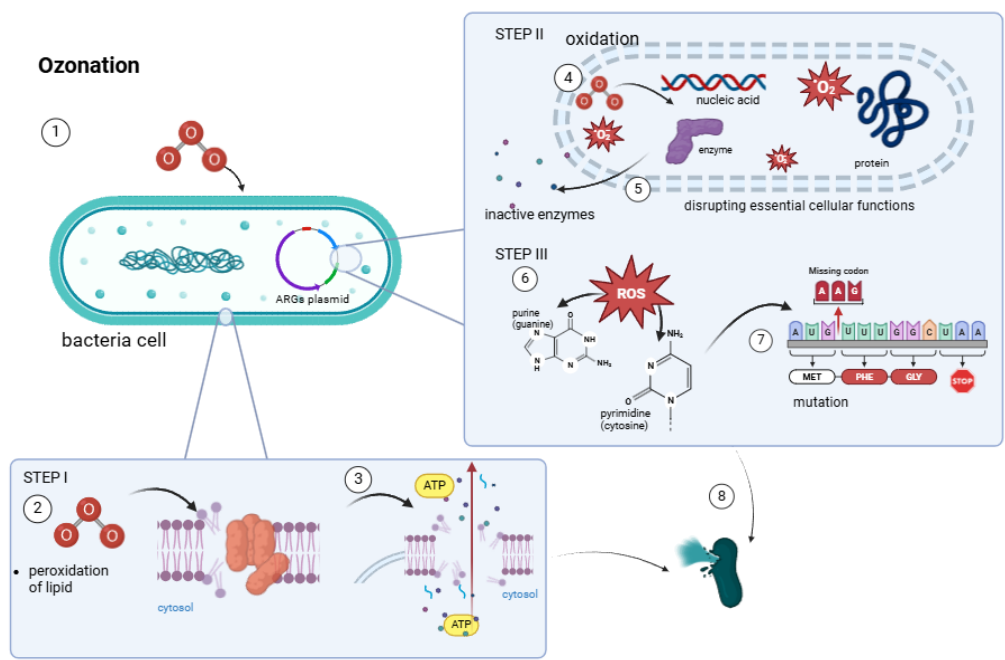

**Figure 4.** Mechanism of ozonation in bacteria cell; 1-use of $O_3$; 2- $O_3$ leads to peroxidation of lipid and destabilization of cell membrane, 3- A damaged membrane is easy way to loss of cellular components e.g. ATP, water and nutrients, 4- (in cytosol) oxidation of nucleic acid, proteins and enzymes 5- Consequence an oxidation process is inactivation enzyme 6- ROS influence on purine and pyrimidine and (7) consequence of this phenomenon are mutation in DNA 8- In the effect a bacterial cell is lysing (created with BioRender https://www.biorender.com/ ) (Bitter et al., 2017; Boševski et al., 2020; Rangel et al., 2022)

In another study, ozonation increased the concentrations of *van*A (4-fold increase) and *bla*$_{VIM}$ (7-fold increase) genes, while decreasing the copy number of the *erm*B gene by 50%, when the concentration of 0.9 ± 0.1 g ozone per 1 g DOC has been used (Alexander et al., 2016). The experiment for removal AR-*E. coli* and AR-*P. aeruginosa* and ARGs by ozonation showed to an acceptable reduction of ARGs, but these genes were still present after exposure to 45 mg $O_3$ L$^{-1}$ for 15 min (Baghal Asghari et al., 2021). According to some researchers, microbubble and nanobubble ozonation can increase ozone stability and enhance the effectiveness of disinfection. Hsiao et al. (2023) found that microbubble ozonation effectively degraded ARGs in hospital wastewater at the applied dose of 132 mg $O_3$ L$^{-1}$ min (15 min response time).

**Table 2.** The effectiveness of modified ozonation treatments in removing ARB and ARGs from hospital wastewater.

| Dose | ARGs/ARB | Reduction | References |
|---|---|---|---|
| **111 mg $O_3$ L$^{-1}$ 34 g h$^{-1}$, t=20 / 40 / 80′** | *Enterobacteriaceae* | 88.1 / 95.1 / 97.2 % | (Azuma et al., 2022) |
| | ESBL *Enterobacteriaceae* | 79.8 / 92.2 / 94.6 % | |
| **111 mg $O_3$ L$^{-1}$ 34 g h$^{-1}$, t=20′** | *tet*G, *bla*$_{GES-1}$ | 100% | |
| **111 mg $O_3$ L$^{-1}$ 34 g h$^{-1}$, t=20 / 40 / 80′** | *Citrobacter* | >99 / >99 / 100% | |
| | *Escherichia* | 100 / >99 / >99% | |
| | *Klebsiella* | >99 / >99 / >99% | |
| | *Acinetobacter* | >99 / >99 / >99% | |
| | *Pseudomonas* | 100 / >94 / >97% | |
| **8.6 mg $O_3$ L$^{-1}$, t=100′** | CRE, VRE, MRSA, MDR *Acinetobacter*, MDR *P. aeruginosa* | 100% | (Azuma et al., 2019) |

| 4 mg $O_3$ $L^{-1}$, pH = 7.0, t=20′ Autoclave time: 15′ with 5.5 mg DOC $L^{-1}$ | *Enterococcus, Staphylococcus, E. coli* | >90% | (Heß & Gallert, 2015) |
|---|---|---|---|
| 1.0 mg min $L^{-1}$, t=10′ | VRE, MRSA, MDR *Acinetobacter*, MDR *P. aeruginosa* | >99.9% | (Azuma & Hayashi, 2021) |
| 45 mg $L^{-1}$, t=15′ | *E. coli* *P. aeruginosa* | >99% | (Baghal Asghari et al., 2021) |
| | *sul*1, *bla*TEM, *bla*CTX, *qnr*S, *bla*VIM | >99% | |

Unfortunately, the ozonation of wastewater can give rise to numerous toxic substances. This is related to the occurrence of impurities in the wastewater (Pulicharla et al., 2020). By-products can include ketones harmful to living organisms and carcinogenic bromines and aldehydes. These contaminants are often difficult to remove and pose a serious threat to the environment and public health (Ikehata, 2019; Pulicharla et al., 2020).

*3.3. Chlorine disinfection*

Chlorination is the most widely used method of wastewater disinfection due to low cost and chlorine's ability to inactivate microorganisms. Chlorine disinfection prevents the spread of pathogens and reduces the prevalence of infections (Bridges et al., 2020; Xu et al., 2018; Zheng et al., 2017). Chlorine compounds such as NaOCl are added to wastewater to eliminate pathogens (H. Wang, Wang, et al., 2020). Free chlorine ($Cl_2$) is the most commonly used form of chlorine for disinfection. However, around 600 chlorine by-products have been identified in wastewater treated with this disinfection method (Richardson & Postigo, 2015). Chlorine compounds exert a negative impact on the environment by disrupting biological and biochemical processes in aquatic organisms (Emmanuel et al., 2004). Chlorine compounds can also penetrate soil to reach groundwater and contaminate sources of drinking water, thus posing a public health threat (Tabernacka, 2014). The mechanism of chlorination was illustrated on figure 5.

Chlorine dioxide ($ClO_2$) is an alternative chlorine compound for ARB elimination from wastewater. This compound is a highly effective biocidal agent, but the preparation and storage of chlorine dioxide standards is difficult and hazardous (Luo et al., 2020).

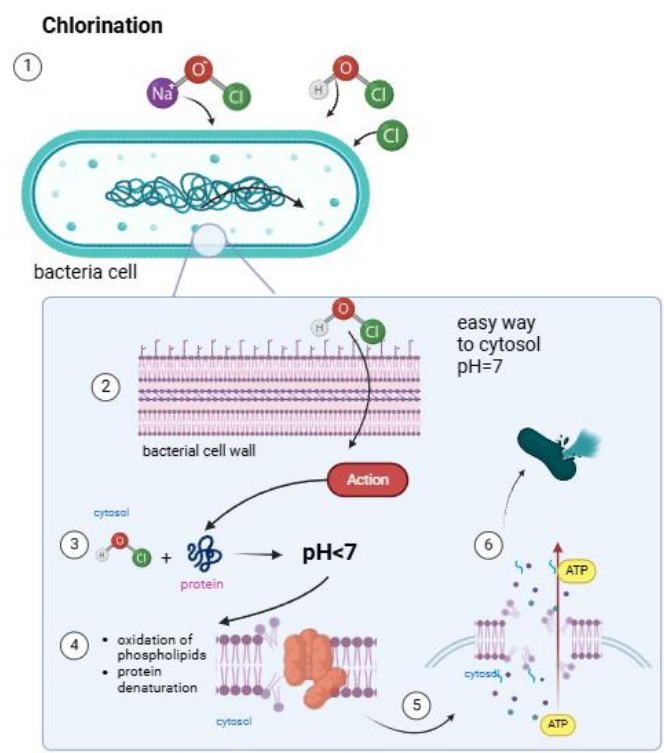

**Figure 5.** Mechanism of chlorination in bacteria cell; 1-use of chlorine compounds; 2- It is easy to run through a cell membrane by HOCl which is neutral molecule, ; 3- (in cytosol) acidification of cytosol by interaction between HClO and protein/amine e. g. HClO is a cause of oxidation -SH group in cysteine; 4- Consequence an acidified cytosol is destabilization of cell membrane by oxidation of phospholipids and protein denaturation, 5- A damaged membrane is easy way to loss of cellular components e.g. ATP, water and nutrients, 6- In the effect a bacterial cell is lysing (created with BioRender https://www.biorender.com/ )(Condon et al., 2005; Muñoz-castellanos et al., 2021)

Chlorine oxidizes organic compounds, including proteins, lipids, and nucleic acids, in microbial cells. Chlorination causes irreversible damage to the bacterial cell membrane and inactivates microorganisms (Bridges et al., 2020; Xu et al., 2018; Zheng et al., 2017). The effectiveness of chlorination in removing ARB and degrading ARGs and antibiotics is determined by the contact time, the chlorine dose, and the applied chlorine compounds (Bridges et al., 2020). Zheng et al. (2017) observed that chlorine did not fully degrade bacterial DNA and that higher HOCl concentrations were required to break bacterial cell walls and completely inactivate ARB (Table 3) (Stange et al., 2019). It should also be noted that chlorine can promote the spread of genetic material via HGT (Jin et al., 2020) because ARG concentrations can increase in wastewater after ARB removal (Liu et al., 2018). Research has also shown that bacterial resistance to antibiotics such as chloramphenicol and ampicillin can increase 5- or 6-fold after chlorination. An NaClO dose of 4 mg $L^{-1}$ induced only minor damage to *P. aeruginosa* cells and contributed to the evolution of a chlorineresistance mechanism in this bacterial species (Hou et al., 2019). The applied chlorine doses have to be regularly monitored. Jin et al. (Jin et al., 2020) found that chlorination intensified HGT, whereas Yuan et al. (2015) reported that HGT was enhanced by low concentrations of chlorine (80 mg min $L^{-1}$) (Yuan et al., 2015). Interestingly, some studies revealed high adaptability of *K. pneumoniae* to chlorine compounds, as the same strain was detected in clinical samples and in hospital wastewater before and after disinfection with active chlorine (0.6 mg $L^{-1}$) (Popa et al., 2021). Similar observations were made by Rolbiecki et al. (2022) who found that the abundance of ARGs did not decrease after chlorination.

**Table 3.** The effectiveness of modified chlorination treatments in removing ARB and ARGs from hospital wastewater.

| Dose | ARGs/ARB | Reduction | References |
|---|---|---|---|
| $Cl_2$ 1-2 mg $L^{-1}$ | *str*B, *tet*A, *tet*B, *aac*C2, *sul*2 | 10-20% | (H. Wang, Wang, et al., 2020) |
| NaClO 0.2 mg $L^{-1}$, t=90′ | *bla*$_{TEM}$, *bla*$_{CTX-M}$, *bla*$_{SHV}$ | No change | (Rolbiecki et al., 2022) |
| $Cl_2$ | *E. coli* | 86,3% | (Bojar et al., 2021) |

*3.4. Other methods of ARB populations' limitation and decreasing of ARGs and antibiotics concentration*

There are also other methods of wastewater disinfection whose effectiveness and minimal negative side effects encourage their wider use. However, despite promising results under laboratory conditions, these methods often fail in practice because it is impossible to fully control the process conditions. Their often high installation or operating costs might cause other problems. The economic aspect remains one of the most important factor when choosing a disinfection method for wastewater.

*3.4.1. Electrocoagulation*

Electrocoagulation (EC) is an electrochemical disinfection technique that is less expensive and more environmentally-friendly than conventional disinfection methods. Mechanism of EC eliminates bacteria by generating $Fe^{3+}/Al^{3+}$ ions, which neutralize the surface charges of bacteria. This is followed by adsorption of the bacteria in the presence of metal hydroxides, forming flocs. The flocs sediment and eventually flotation occurs (Boudjema et al., 2024; Harif et al., 2012). The effectiveness of EC disinfection is largely dependent on environmental pH. As one study show the ARGs (*sul*1, *sul*2, *tet*O, and *tet*X) present in the secondary effluent of a WWTP were effectively removed using EC at a current density of 20.0 mA/cm² under neutral pH conditions. The primary removal mechanism involved the adsorption and entrapment of ARGs within precipitated flocs. Higher current densities improved the reduction of ARGs, while acidic and neutral pH conditions were favorable for their removal via EC. Additionally, pretreatment with conventional UV disinfection enhanced the efficiency of ARG removal in the subsequent EC process (L. Chen, Xu, et al., 2020). The presence of insoluble iron (II) and (III) hydroxides and high current density promote the formation of hydroxide crystals and increase the efficiency of disinfection (L. Chen, Xu, et al., 2020; Gomes et al., 2019; Mosaka et al., 2023a). Electrocoagulation is a potentially effective disinfection method, but further research is needed to validate its applicability in wastewater treatment.

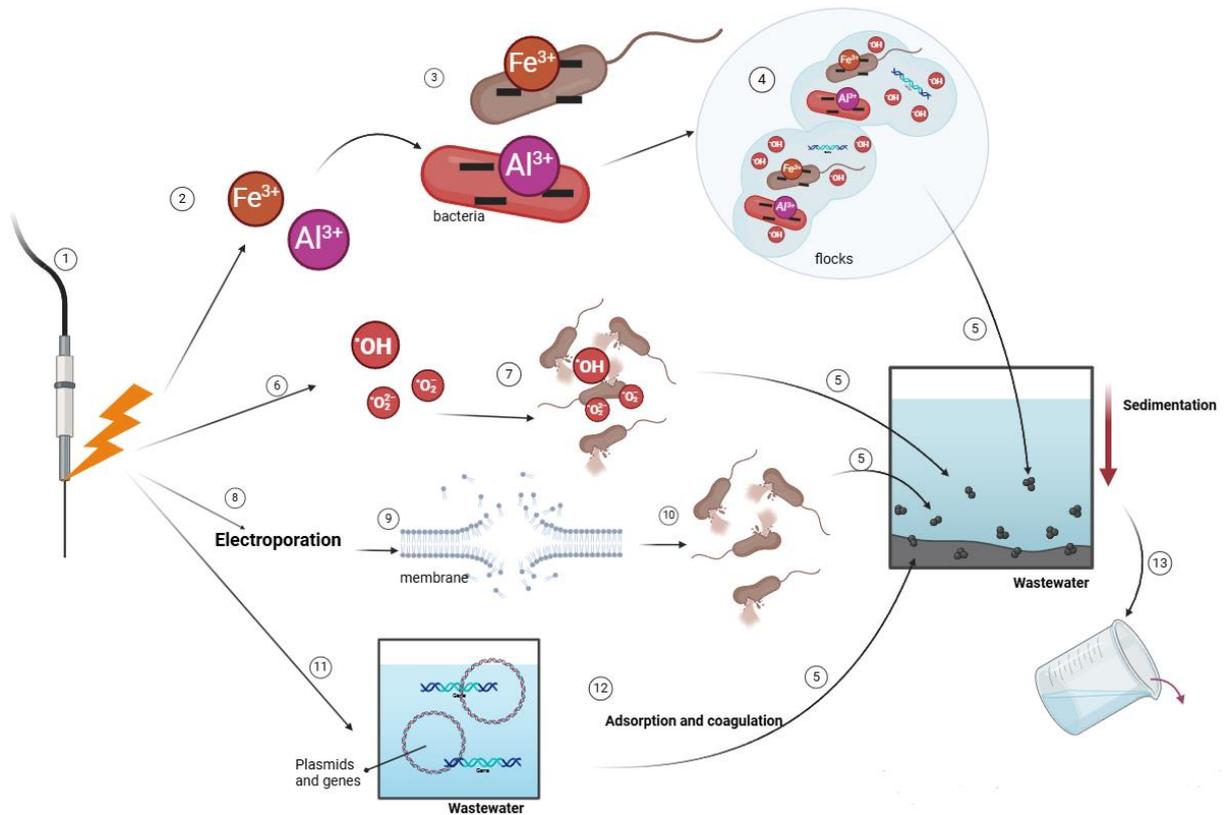

**Figure 6.** Mechanism of electrocoagulation in bacteria cell; 1, 2- formation of $Fe^{3+}$ or $Al^{3+}$ on electrodes under voltage; 3 - neutralization of negative charges on the bacterial surface; 4 - creation of metal hydroxides (iron and aluminum) that adsorb to bacteria and ARGs, as a result of which flocs are formed; 5 - sedimentation/formation of precipitates; 6 - formation of ROS that damage the cell membrane; 7 - bacterial lysis; 8 - electroporation; 9 - destabilization of the cell membrane; 10 - bacterial lysis; 11, 12 - adsorption and coagulation of genetic material, 13 - flotation (created with BioRender https://www.biorender.com/ ) (Boudjema et al., 2024; Delaire et al., 2016; Harif et al., 2012; Holt et al., 2006)

### 3.4.2. Photocatalysis

Photocatalysis is a wastewater treatment method that relies on UV or visible light and catalysts. In this process, a photocatalyst is activated by light to trigger photocatalytic reactions that degrade water pollutants (Ghernaout & Elboughdiri, 2020). Titanium dioxide ($TiO_2$) is most commonly used as a photocatalyst (Moreira et al., 2018). Photocatalysis removed a high percentage of sul1 (99.71%), *int*I1 (98.36%) and *bla*TEM (97.41%) genes, on the other hand the lowest deletion percentage of a gene was observed for *sul*2 gene (82.49%) . Other study showed ARB was removed at a lower level in comparison to ARGs from hospital wastewater (>80%) (Kaliakatsos et al., 2023).

### 3.4.3. Peracetic acid

Some disinfection chemicals have a broad spectrum of activity. This group of compounds includes PAA which is a colorless mixture of hydrogen peroxide and acetic acid (Kwalska, 2016; Luukkonen et al., 2014). Research has shown that PAA is more effective in inhibiting the formation of mutagenic and toxic compounds than chlorine (Hassaballah et al., 2020).

Peracetic acid oxidizes sulfur and sulfhydryl bonds in bacterial proteins, increases the permeability of bacterial cell walls, and inhibits intracellular transport (Eramo et al., 2017; Kitis, 2004). This compound effectively removes ciprofloxacin-resistant bacteria. Its disinfectant properties are enhanced during prolonged contact with organic matter and at higher concentrations (Chhetri et al., 2022). In a study by Campo et al. (2020), PAA effectively reduced ARB concentrations in municipal wastewater (Table 4). After 16 min of exposure to the PAA dose of 3 mg L$^{-1}$, the percentage of ampicillin-resistant bacteria in the overall bacterial population was reduced from 40% to 2.7%. Peracetic acid was less effective in degrading ARGs. A 0.4 log reduction was achieved in the concentration of the *sul*2 gene, whereas the concentrations of sul1 and *tet*A genes remained unchanged after 30 min of exposure to the PAA dose of 1 mg L$^{-1}$ (Luprano et al., 2016).

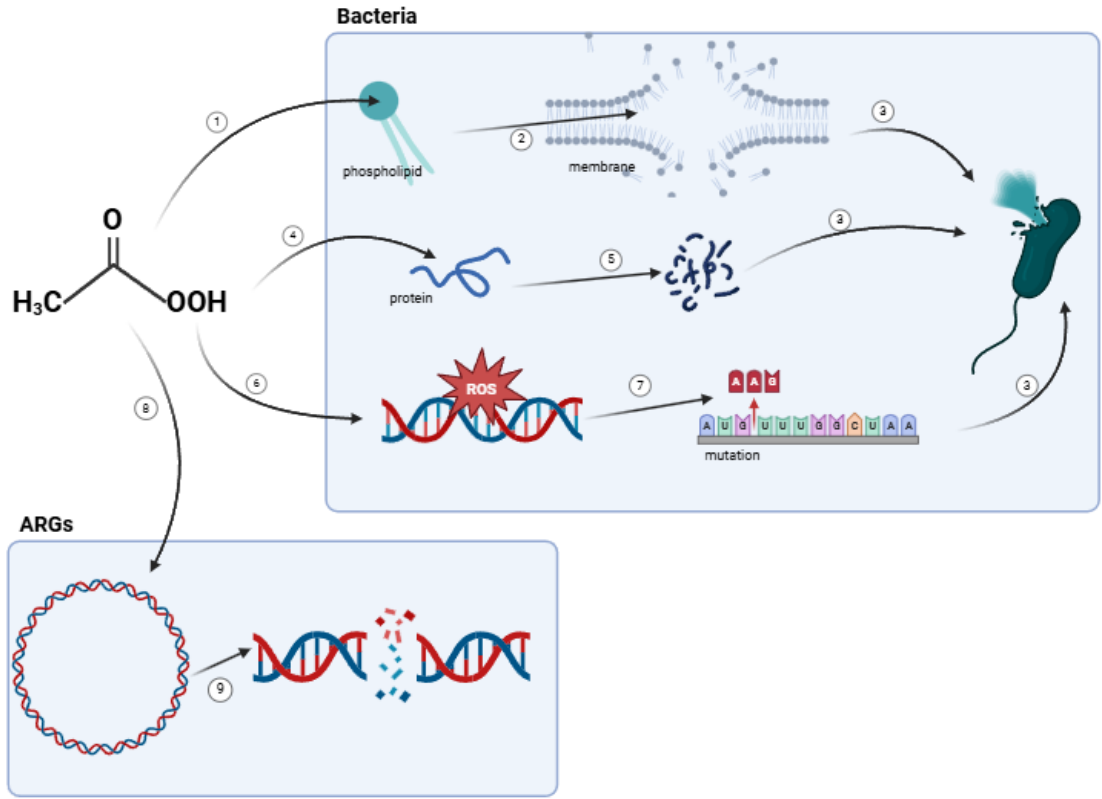

**Figure 7.** Mechanism of peracetic acid action in bacteria cell; 1- PAA reaction with membrane lipids; 2 - increase in permeability of membrane and its integrity loss; 3 - bacterial lysis; 4 - oxidation of -SH groups; 5 - denaturation of proteins and inactivation of enzymes; 6 - generation of ROS acting on gDNA; 7 - mutation within genes; 8 - disruption of phosphodiester bonds; 9 - DNA fragmentation (created with BioRender https://www.biorender.com/ ) (Leggett et al., 2016; Viola et al., 2018; D. Wang et al., 2020; T. Zhang et al., 2020)

The counts of MDR *E. coli* were more effectively reduced by the PAA dose of 4 mg L$^{-1}$ (7 min) than the NaOCl dose of 2 mg L$^{-1}$ (4 min) (Balachandran et al., 2021). However, some researchers have found that low doses of PAA and short contact times can increase ARB concentrations even 40 fold in municipal wastewater, which gives serious cause for concern (J. J. Huang et al., 2013). Peracetic acid contains organic acetic acid (CH$_3$COOH) which can stimulate bacterial growth (Kitis, 2004). However, the applicability of PAA for eliminating ARB and ARGs from hospital wastewater remains insufficiently investigated.

*3.4.4. Advanced oxidation processes*
To overcome the limitations of conventional disinfection methods, various modifications have been proposed to increase the effectiveness of ARGs and ARB removal. Advanced oxidation processes (AOPs) combine popular disinfection techniques, such as UV radiation and ozonation, with substances that promote the elimination of ARGs and ARB. Advanced oxidation is one of the most popular and cost-effective methods of removing organic and chemical substances from wastewater (Calcio Gaudino et al., 2021). Hydrogen peroxide (H$_2$O$_2$) and TiO$_2$ are frequently used in AOPs to catalyze pollutants. UV/LED-assisted AOPs and the photo-Fenton (PF) process are also effective wastewater treatment methods. The main advantage of advanced oxidation is that these processes do not produce toxic substances (Y. di Chen et al., 2021; Park et al., 2016; Xiao et al., 2019). Advanced oxidation processes generate ROS which damage cell membranes, proteins, and genetic material in bacterial cells. They produce lipid peroxidation, which disintegrates cell membranes and increases permeability. ROS also have an effect on proteins, modifying their structure and function and interfering with essential bacterial life processes. Furthermore, ROS disrupt the cell wall's peptidoglycan, rendering it more vulnerable to injury. At the genomic level, they alter nitrogenous bases, inhibiting DNA replication and transcription and causing single- and double-strand breaks. As a result, the genetic material is fragmented, and it triggers cell death (Hong et al., 2019; C. Yang et al., 2020). Reactive oxygen species can be divided into radicals and molecules without unpaired electrons, such as singlet oxygen, ozone, and hydrogen peroxide (Y. di Chen et al., 2021; Dutta et al., 2019). Advanced oxidation processes promote the degradation of ARGs in wastewater treatment (Zhou et al., 2020). In addition, the effective UV dose can be lowered in the presence of ROS (H. Wang, Wang, et al., 2020). The effectiveness of AOPs in removing ARGs and ARB from hospital wastewater is presented in Table 4.

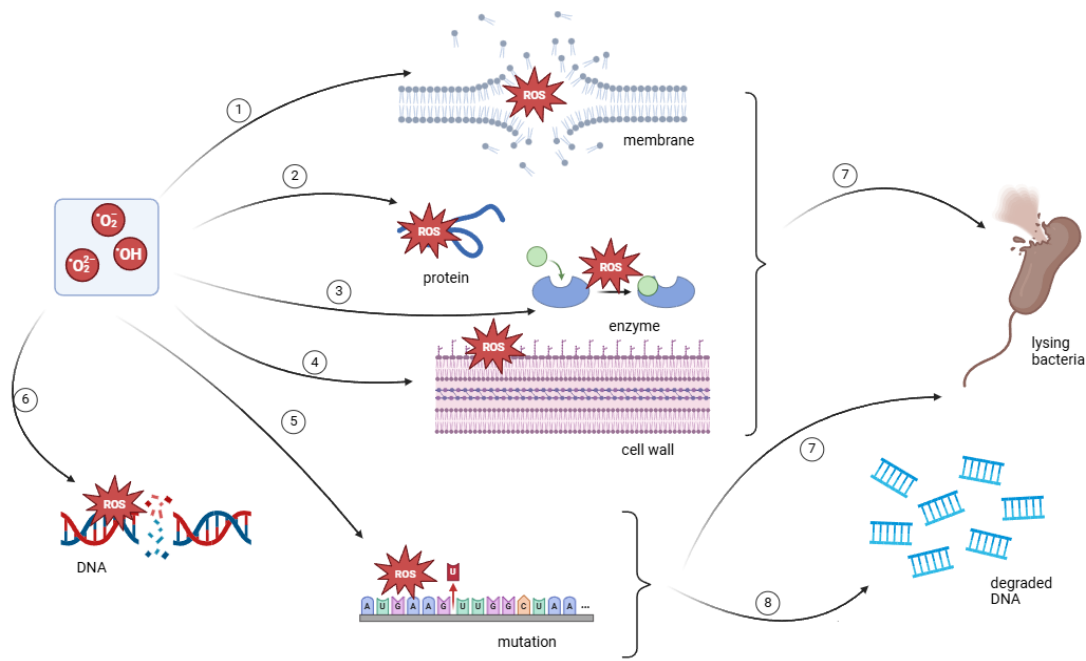

**Figure 8.** Mechanism of ROS action in bacteria cell; 1 - damage of the cell's membrane by ROS activity - lipid peroxidation; 2 - oxidation of the -SH group is leading to changes in protein structures and function; 3 - reduced enzyme activity - disruption of basic bacterial life processes; 4 - fragmentation of peptidoglycan causing more cell permeability; 5 - mutations of nitrogenous principles - problems in replication and transcription; 6 - rupture of single and double strands of DNA; 7 – cell's death; 8 - DNA fragmentation (created with BioRender https://www.biorender.com/ )(Hong et al., 2019; C. Yang et al., 2020)

- *UV/$H_2O_2$*

Despite the fact that disinfection methods involving UV radiation and $H_2O_2$ have been extensively investigated, there is no conclusive evidence that these techniques effectively remove micropollutants (Umar, 2022; Umar et al., 2019). In some studies, $H_2O_2$ disinfection inactivated bacteria of various genera, and higher doses of $H_2O_2$ contributed to reducing the abundance of ARGs. At the same time, higher doses of $H_2O_2$ promoted the degradation of antibiotics (Beretsou et al., 2020). However, other researchers found that $H_2O_2$ disinfection decreased ARG concentrations by up to 2.9 log (Umar, 2022). Ferro et al. (2016) reported that UV/$H_2O_2$ disinfection not only eliminated *E. coli* from wastewater, but also decreased the copy numbers of *qnr*S and *bla*TEM genes, whereas other carbapenem-resistance genes were not successfully degraded. UV radiation and $H_2O_2$ trigger the production of ROS that effectively damage ARB and ARGs (Giannakis et al., 2017; Umar et al., 2019). However, the efficiency of combined UV/$H_2O_2$ disinfection is difficult to evaluate because these methods differ in their ability to degrade ARGs due to the fact that longer amplicons are more prone to degradation.

- *UV/chlorine*

Ultraviolet light has a marginal impact on the environment, and this treatment method can be effectively combined with chlorination. Despite its potent bactericidal activity, chlorine promotes the production of harmful substances that can increase AR in the environment (D. Li & Gu, 2019; L. Wang et al., 2019). For this reason, chlorination can be combined with UV radiation to decrease disinfection costs and minimize the spread of toxic disinfection by-products to the environment, which is a major concern when chlorine is applied alone. Zhang et al. (2015) revealed that combined UV/$Cl_2$ disinfection (optimal parameters: 62.4 mJ UV $cm^{-2}$ and 25 mg $Cl_2$ $L^{-1}$) was more effective than UV and $Cl_2$ treatments applied alone. The largest reduction occurred in the case of genes *tet*X and *16S* rRNA. Moreover, *sul*1, *tet*G and *int*I1 genes were also sensitive to these processes (Table 1, Table 2).

- *UV/$O_3$*

Combined UV and ozone disinfection involves UV light with a wavelength shorter than 300 nm because ozone intensifies the sterilizing properties of UV radiation (Bracamontes-Ruelas et al., 2022; Cuerda-Correa et al., 2020). Reactive oxygen species play an important role in UV/$O_3$ treatment by oxidizing organic matter (Y. di Chen et al., 2021; Rekhate & Srivastava, 2020). However, this disinfection method has not been extensively studied (Igere et al., 2020). According to Jäger et al. (2018), UV/$O_3$ treatment decreased the concentrations of ARGs (*ecfx*, *ycc*T, *mec*A, *bla*$_{CTXM32}$, *erm*B, *bla*$_{TEM}$, *sul*1, *int*I1) in wastewater by more than 90%.

- *UV/$S_2O_8^{2-}$*

UV disinfection with persulfate (PS) is an advanced wastewater treatment method that eliminates microorganisms. This process effectively combines the disinfecting properties of UV radiation and PS (Fu et al., 2019; Q. Zhang et al., 2015). Numerous research studies have demonstrated that combined UV/PS disinfection is more effective than UV radiation alone. UV-C-driven disinfection with PS was 59% more effective than the UV-A-assisted PF process. Zhou et al. (2020) also found that combined UV/PS disinfection induced a greater and more rapid decrease in the copy numbers of ARGs than UV alone, and the total reduction in the abundance of ARGs (*int*I2, *int*I2, *qnr*S, *tet*O, *sul*2, *sul*1, *erm*B) was by 0.56 log higher. In addition, UV/PS treatment also eliminated bacteria resistant to UV light. In this method, the removal of MGEs exceeded 76% (Arslan-Alaton et al., 2021; Zhou et al., 2020). The effectiveness of UV/PS treatment in hospital wastewater disinfection has not been sufficiently analyzed and needs to be investigated.

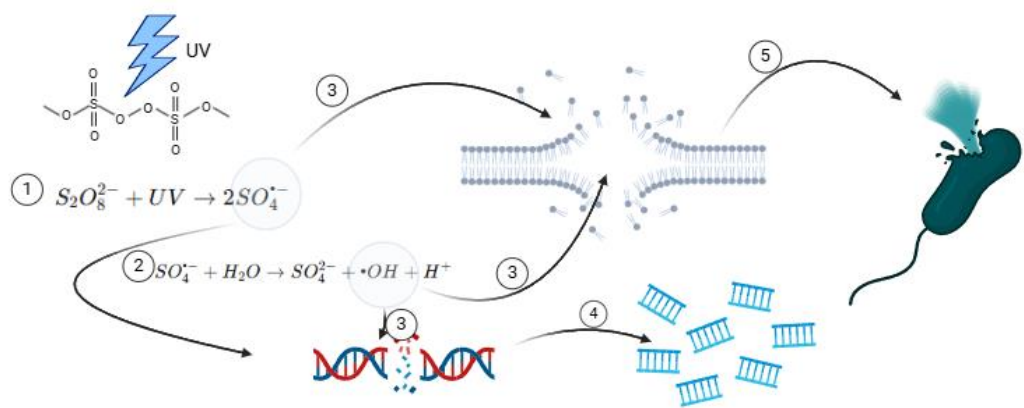

**Figure 9.** Mechanism of persulfate action in bacteria cell; 1 - decomposition of PS to sulfate radicals under the influence of UV, according to the scheme; 2 - generation of hydroxyl radicals due to the reaction of sulfate radicals with water; 3 - peroxidation of phospholipids, denaturation of proteins and oxidation of peptidoglycan under the influence of sulfate and hydroxyl radicals, resulting in increased membrane

permeability; 4 - damaging of nitrogenous bases and deoxyribose due to the influence of hydroxyl and sulfate radicals; 5 - cell death; 6 - fragmentation of genetic material (created with BioRender https://www.biorender.com/ ) (Z. Huang et al., 2023; T. Zhang et al., 2020)

- *Photo-Fenton process*

Photo-Fenton process The PF process is an advanced wastewater treatment technique that was invented by Fenton in 1984 (Akbari et al., 2021; Bracamontes-Ruelas et al., 2022). The PF method combines photocatalysis with the Fenton reaction (Bracamontes-Ruelas et al., 2022), and it is applied to remove organic wastewater compounds that are toxic and not readily biodegradable. In this process, pollutants are oxidized and degraded under exposure to UV/solar radiation in the presence of a catalyst. The following reactions take place simultaneously or successively in the PF process: photocatalysis, Fenton reaction, oxidation, and degradation. In the first stage, UV light or sunlight is used to activate the photocatalyst, usually $TiO_2$ or $Fe^{3+}$ (Bracamontes-Ruelas et al., 2022). Under exposure to UV radiation, the photocatalyst begins to generate electron-hole pairs. Small quantities of hydrogen peroxide ($H_2O_2$) and iron ($Fe^{2+}$ or $Fe^{3+}$) catalysts are used in the Fenton reaction (Akbari et al., 2021; Y. di Chen et al., 2021; L. Chen, Zhou, et al., 2020). Hydrogen peroxide is decomposed in the presence of iron, and it produces hydroxyl radicals (˙OH) which are powerful oxidants (Cuerda-Correa et al., 2020). Hydroxyl radicals react with organic pollutants in wastewater. Radicals decompose organic compounds into simpler products that are less toxic and easier to degrade, and they effectively target extracellular ARGs in wastewater. One of the greatest advantages of the PF process is that it is more environmentally-friendly than other advanced treatment methods because it relies on natural sources of energy (sunlight) and does not require additional chemicals to degrade highly toxic compounds in wastewater (Akbari et al., 2021; Y. di Chen et al., 2021; Giannakis et al., 2017). However, its effectiveness is influenced by many factors, including the type of the photocatalyst, the intensity of UV or solar radiation, hydrogen peroxide and iron concentrations, and the composition of wastewater pollutants. Therefore, the PF process has to be adapted to local conditions and the type of pollutants in treated wastewater. The Fenton reaction occurs within a temperature range of 20-40°C, and it slows down at 40-50°C because $H_2O_2$ is broken down into water and oxygen (Cuerda-Correa et al., 2020). The solar PF process is a variation of the PF method that relies on sunlight to launch the process (Cuerda-Correa et al., 2020).

The phenomenon of PF is marked by its high efficiency in the degradation of ARB and ARGs. This efficiency is attributed to the generation of hydroxyl radicals, which possess a high oxidation potential of approximately 2.8 volts (V) (Mosaka, 2023b). Furthermore, solar-PF utilizes a natural source of radiation during operation, thereby significantly reducing the cost of energy consumed during the process. Solar radiation, a viable source of energy, facilitates the large-scale implementation of this technology in countries that exposure high levels of sunshine throughout the year (Sharma et al., 2023). A notable advantage of this technology is that it does not necessitate a sophisticated infrastructure (Fenton, 2017). Photoreactors are straightforward and cost-effective to implement, representing a viable alternative to other technologies, such as ozonation, cold plasma, or UV-C radiation. A comparison of solar-PF with chlorination reveals a discrepancy in the degree of negative environmental impact (Foteins et al. 2018).

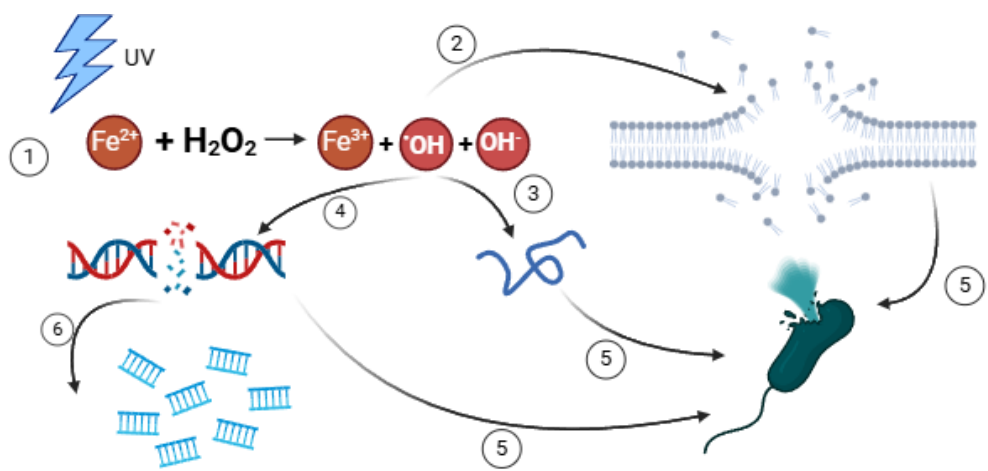

**Figure 10.** Mechanism of Photo-Fenton process in bacteria cell; 1- in the presence of sunlight or UV light, iron (II) ions initiate the Fenton reaction, which proceeds according to the scheme; hydroxyl radicals damage - 2 - cell membranes; 3 - proteins; 4 - DNA; 5 - which leads to cell death; 4, 6 - hydroxyl radicals also affect and damage extracellular genetic material, leading to its fragmentation (created with BioRender https://www.biorender.com/ ) (Giannakis et al., 2018)

**Table 4.** The effectiveness of AOPs in removing ARB and ARGs from hospital wastewater.

| Method | Dose | ARB/ARGs | Reduction | References |
|---|---|---|---|---|
| **PAA** | 50 mg L$^{-1}$, t=3′ | ciprofloxacin-RB | 99.99% | (Chhetri et al., 2022) |
| **UV/PAA** | T=3min | *E. coli* NDM-1 | 100% | (Li et al., 2024) |
| **UV+chlorine** | UV 320 mJ cm$^{-2}$ Cl$_2$: 1 mg L$^{-1}$ | *aac*C2, *sul*II, *str*B, *tet*A, *tet*B AR *Morganella morganii* and *E. faecalis* | 50-60% | (H. Wang, Wang, et al., 2020) |
| | UV 320 mJ cm$^{-2}$ Cl$_2$: 2 mg L$^{-1}$ | *aac*C2, *sul*II, *str*B, *tet*A, *tet*B AR *Morganella morganii* and *E. faecalis* | 70-80% | (H. Wang, Wang, et al., 2020) |
| | UV 62.4 mJ cm$^{-2}$ Cl$_2$: 25 mg L$^{-1}$ | | | |
| **UV-LED** | Λ=265nm | *tet*A | <100% | (Zhao et al., 2023) |
| | | *kat*1 | <100% | |
| | | *amp*C | <100% | |
| | Λ=280 nm | T=11,99 min 9,6 mJ cm$^{-2}$ — CRE | 99% | (Azuma et al., 2024) |
| | | T= 1,3 min 31,9 mJ cm$^{-2}$ — ESBL-E | | |
| | | T=1,5 min 32,8 mJ cm$^{-2}$ — MDRA | | |
| | | T= 1,5 min 29,3 mJ cm$^{-2}$ — MDRP | | |
| | | T=0,8 min — MRSA | | |

|  |  | 70,6 mJ cm$^{-2}$ |  |  |
|---|---|---|---|---|
|  |  | T= 3,5 min<br>10,4 mJ cm$^{-2}$ | VRE |  |
|  |  | T=1,5 min<br>40,8 mJ cm$^{-2}$ | *Acinetobacter* |  |
|  |  | T= 3,5 min<br>12,7 mJ cm$^{-2}$ | *Enterococcus* |  |
|  |  | T= 2,5 min<br>16,7 mJ cm$^{-2}$ | *E. coli* |  |
|  |  | T= 1,9 min<br>14,1 mJ cm$^{-2}$ | *P. aeruginosa* |  |
|  |  | T= 1,5 min<br>28,3 mJ cm$^{-2}$ | *S. aureus* |  |
| **UV/H$_2$O$_2$** | 340, 1700, 3400 mg/L |  | *amp*C, *mec*A | >90% | (Umar, 2022) |
| **UV + EC** |  |  | *sul* gene | 100% | (L. Chen, Xu, et al., 2020) |
|  |  |  | *tet* gene | 97% |  |
| **UV+ ozonation** |  |  | *erm*B | 98% | (Jäger et al., 2018) |
|  |  |  | *sul* | 96% |  |
|  |  |  | *bla*$_{SHV/TEM}$ | 91% |  |

- *Nanoparticles*

Nanoparticles are ultrafine particles with a diameter of 1-100 nanometers (nm), and they have numerous practical applications, for instance in wastewater treatment (Balarak & Mostafapour, 2019). Nanoparticles enhance the removal of pollutants and microorganisms from wastewater, and they can increase the efficiency of treatment processes. Nanoparticles can be used for the adsorption and binding of various types of contaminants, including heavy metals, organic substances, chemical compounds, and microplastics (Malakootian et al., 2019; Ren et al., 2018). Nanoparticles with a large specific surface area, such as carbon nanotubes, graphene and metallic nanoparticles, effectively adsorb penicillin G (Chavoshan et al., 2020). Some study showed silica nanoparticles remove Penicillin G. The highest removal effect has occurred in pH=7 and decreasing Penicillin G concentration, which caused the increase of the adsorption capacity (Masoudi et al., 2019). Silver nanoparticles have antibacterial and antifungal properties, and they can be used in wastewater disinfection. Nanosilver is particularly effective in eliminating bacteria and other microorganisms. The mechanism of the silver nanoparticle's action against ARB involves several steps. Initially, the nanoparticles bind to bacterial DNA bases, subsequently disrupting membrane function. Additionally, once internalized within the bacterial cell, the nanoparticles cause cellular damage. The degradation of silver nanoparticles releases silver ions, which interact with thiol-containing proteins in the bacterial cell wall, impairing essential cellular functions (Ezeuko et al., 2021a). Nanoparticles have a growing number of applications, but their effectiveness in wastewater treatment has not been sufficiently investigated.

*3.4.6. Ionizing radiation*

Ionizing radiation is a process that damages or inactivates microorganisms such as bacteria, viruses and pathogens, and it is used in wastewater treatment to eliminate public health risks. Ionizing radiation has a similar physicochemical mechanism to UV radiation (Ezeuko et al., 2021b). The study demonstrated that even a relatively low dose of gamma radiation (10 kGy) decreased the copy numbers of ARGs in range of 10.8% to 92.6%. Higher radiation doses of 25 kGy and 50 kGy reduced the copy numbers of ARGs by 85.6–98.9% and 96.5–99.8%, respectively (Chu et al., 2021).

# 4. Unintended consequences of disinfection methods for degrading antibiotics and drug residues in hospital wastewater

The widespread use of antibiotics in medicine and agriculture has contributed to the spread of these substances in the natural environment (Proia et al., 2018). Pollution with antibiotics and drug residues poses a significant challenge for public health and water ecosystems (Reina et al., 2018). The discussed pre-treatment methods not only decrease the concentrations of ARB and ARGs, but they also effectively degrade antibiotics in whole or in partially. Therefore, most disinfection techniques remove microorganisms and degrades antibiotics to their TPs as a part of daily practice. The presence of antimicrobials in wastewater can contribute to the spread of AR, which poses a serious threat to public health. Antibiotics and their transformation products (TPs) can induce AR in environmental bacteria (Proia et al., 2018). Therefore, these substances have to be eliminated from wastewater. The effectiveness of antibiotic degradation is determined by the applied method, as well as the type and structure of antibiotics (Gomes et al., 2019; Serna-Galvis, Berrio-Perlaza, et al., 2017).

*4.1. UV*

Antibiotics can be degrade from wastewater with the use of UV radiation. UV light degrades antimicrobials by cleaving their chemical bonds. In this process, antibiotics like meropenem and imipenem are decomposed into simpler and less toxic chemical compounds (Reina et al., 2018; Rizzo et al., 2020). The effectiveness of UV treatment in degrading antibiotics is affected by the UV dose, i.e. the amount of UV energy applied to water or wastewater. High UV doses are required to effectively degrade antibiotics (norfloxacin, ciprofloxacin and levofloxacin) and other persistent pollutants (Geng et al., 2020). Wastewater treated with UV light should be regularly monitored to ensure that antibiotics like amoxicillin, ciprofloxacin are effectively eliminated and to adjust the UV dose if necessary (Rizzo et al., 2013).

*4.2. Chlorination*

Chlorination could lead to degrade antibiotics and drug residues in wastewater. Chlorine is a strong oxidant that reacts with organic compounds such as antimicrobials. These reactions degrade antibiotics (penicillins, cephalosporins and fluoroquinolones) and reduce their concentrations in wastewater. Notably, it eliminated over 90% of the antibiotic concentration within just 20 minutes (Serna-Galvis, Ferraro, et al., 2017). Chlorination can also decrease the biological activity of antimicrobials, thus rendering them less toxic for living organisms and the environment (Yuan et al., 2015). Chlorination may be effective in degrading some antibiotics, but it does not fully eliminate all pharmaceuticals from wastewater (Serna-Galvis, Ferraro, et al., 2017). The effectiveness of chlorination in removing antibiotics from hospital wastewater is presented in the Table 5.

**Table 5.** The effectiveness of modified chlorination treatments in removing antibiotics from hospital wastewater.

| Method | Conditions | Antibiotic | Reduction | References |
|---|---|---|---|---|
| **HOCl** | t=20′<br>0.55 mol/L Cl$^-$ | Cephalexin<br>Cephadroxyl<br>Cloxacillin<br>Oxacillin<br>Ciprofloxacin<br>Norfloxacin | 90% | (Serna-Galvis, Berrio-Perlaza, et al., 2017) |
| **Cl$_2$** | 1.0 mM Cl$_2$ | Chloramphenicol<br>Ciprofloxacin<br>Sulfamerazine | 50% | (H. Wang, Shi, et al., 2020) |
| **Electrogenerated active chlorine** | t=40′ | Fluoroquinolones | 100% | (Serna-Galvis, Jojoa-Sierra, et al., 2017) |

*4.3. Ozonation*

Ozonation is an advanced wastewater disinfection method that can effectively degrade antibiotics (C. Wang et al., 2020). Ozone decomposes antimicrobials through oxidation (Wajahat et al., 2019). Ozone is a potent oxidant that reacts with organic pollutants, including antibiotics, by cleaving their chemical bonds. These reactions can break down antibiotics into less toxic products (Anthony et al., 2019; Gomes et al., 2019; Wallmann et al., 2021). The effectiveness of ozonation (Table 6) is determined by many factors, including the ozone concentration, the contact time, the environmental pH, the type of antibiotic, and its concentration in wastewater. In some cases, ozonation can significantly reduce antimicrobial levels (Gorito et al., 2022). Ozonation decomposes antibiotics into substances that are less toxic or more persistent than the parent compound. These products have to be monitored to ensure that they do not pose a new threat to the environment (Gomes et al., 2019). Aleksić et al. (2021) showed that both methods $H_2O_2/O_3$ are able to stimulate the removal of antibiotics e. g. amoxicillin and ciprofloxacin.

**Table 6.** The effectiveness of modified ozonation treatments in removing antibiotics from hospital wastewater.

| Method | Conditions | Antibiotic | Reduction | References |
|---|---|---|---|---|
| $TiO_2/O_3$ | t= 240′, Flow rate=77 L/h, Light inten. = 38 W.m$^{-2}$, Initial concent. =30 mg/L | Flumequine | *D=52% M=42% | (Lou et al., 2017) |
| $ZnO/O_3$ | 1.38 mg/s, pH=7; t= 35′ | Oxytetracycline | 94% | (Mohsin & Mohammed, 2021) |
| $H_2O_2/O_3$ | 100 mg L$^{-1}$, t= 30′, 120′ | Amoxicillin Ciprofloxacin | 99% 96% | (Aleksić et al., 2021) |
| $O_3$ | 10 mg $O_3$ L$^{-1}$ min$^{-1}$; 10 min | Fluoroquinolone | 84% | (Rodrigues-Silva et al., 2019) |

* D -degradation, M – mineralization

*4.4. Other methods of antibiotics degradation in hospital wastewater*

*4.4.1. Advanced oxidation processes*

UV radiation is equally effective in eliminating ARGs, and ARB from wastewater. Free radicals generated during AOP could cause the degradation of antibiotics in wastewater as a side effect. Pollutants are degraded by direct UV photolysis or indirect UV photolysis that involves HO* and $SO_2$* (Mao et al., 2015). UV treatment is less effective in highly contaminated wastewater because organic matter is able to capture UV light and free radicals (Gonz et al., 2023). Balarak and Mostafapour used nickel(II) oxide (NiO) to degrade amoxicillin in pharmaceutical wastewater and reported 96% removal efficiency. They also found that longer exposure to UV radiation increased the rate of amoxicillin degradation and that the removal efficiency was influenced by the concentration of NiO (Balarak & Mostafapour, 2019).

**Table 7.** The effectiveness of modified UV radiation treatments in removing antibiotics from hospital wastewater.

| Method | Conditions | Antibiotic | Reduction | References |
|---|---|---|---|---|
| UV/ZnO | ZnO 10 mg L$^{-1}$, pH 5, ZnO NP 0.1 g L$^{-1}$, t=180′ | Penicillin G | 74.65% | (Chavoshan et al., 2020) |

| UV/PAA | T=8 min | Ampicilin | 96,7% | (Li et al., 2024) |
|---|---|---|---|---|
| UV/NiO | NiO 0.2 g/L<br>AMO 25 mg/L<br>UV= 15 W | Amoxicillin | 96% | (Balarak & Mostafapour, 2019) |
| UV/PS | UV-254 nm, t=1 h<br>[PS] = 0.5 mM<br>dose =330 J | Chloramphenicol | 65% | (Ghauch et al., 2017) |
| $TiO_2$/UV/ $H_2O_2$/ $O_3$ | T= 240′<br>Flow rate=77 L/h,<br>Light intensity=38<br>W.m$^{-2}$, Initial<br>concentration=30<br>mg/L | Flumequine | *D=94%<br>M=76% | (Lou et al., 2017) |
| $TiO_2$/UV/ $H_2O_2$ | | | D=93%<br>M=72% | |
| $TiO_2$/UV/ $O_3$ | | | D=92%<br>M=68% | |
| $TiO_2$/UV | | | D=90%<br>M=62% | |
| UV/$O_3$ | | | D=40%<br>M=10% | |
| | | Sulfamethoxazole | > 99% | |

\* D - degradation, M – mineralization

### 4.4.2. Electrocoagulation

Electrocoagulation (EC) has attracted considerable research interest due to its low cost, simplicity and effectiveness. Research has shown that EC effectively degrades cefazolin, but electrode and energy consumption may be higher in turbid wastewater containing $Al(OH)_3$ flocs, which increases treatment costs (Esfandyari et al., 2019). In another study, amoxicillin with an initial concentration of 54.66 mg L$^{-1}$ was degraded from wastewater in 90.56% after 30 min of exposure to an EC treatment with a current density of 2.31 mA cm$^{-2}$ (Mehrabankhahi et al., 2023).

### 4.4.3. Sulfate radicals

Advanced oxidation processes are chemical treatments that decompose harmful compounds into simple and biodegradable products. Sulfate radical ($SO_4^-$)-based AOPs have emerged as a promising alternative for degrading antibiotics from wastewater. *Sul*fate radicals are generated when compounds such as PS and peroxymonosulfate are thermally activated under exposure to heat, UV or activated carbon. Pirsaheb et al. (Pirsaheb et al., 2020) reported that ciprofloxacin (1 mg L$^{-1}$) and amoxicillin (1 mg L$^{-1}$) were degraded in 99.9% and 99.26%, respectively, after 60 min of exposure to the PS dose of 10 mg L$^{-1}$.

### 4.4.4. Cold plasma

Cold plasma has strong oxidizing properties, and it can be used to degrade antibiotics from wastewater. Cold plasma treatment led to nearly complete degradation of antibiotics. Under optimal conditions (initial pH of 10, electrode spacing of 10 mm, reaction duration of 15 minutes, and an applied voltage of 30 kV), the removal efficiencies for all antibiotics were significant, exceeding 72% for ciprofloxacin and cefuroxime, and surpassing 99% for amoxicillin and ofloxacin, along with the complete elimination of COD and ammonia (Nguyen et al., 2021).

Cold plasma is a method that removes pharmaceutical contaminants from wastewater with high efficiency by generating ROS and RNS (reactive nitrogen species). The following chemicals are of particular interest: •NO, •NO$_2$, and ONOO$^-$ (Gonçalves et al., 2025). Cold plasma, in contrast to ozonation and chlorination, does not necessitate direct contact with the liquid. Furthermore, this method does not necessitate the utilization of supplementary reactants (H$_2$O$_2$, Fe$^{3+}$), in opposite to the approach employed in PF (Gonçalves et al., 2025; Nguyen et al., 2021). Additionally, it features a brief operational time span, ranging from seconds to minutes. This property reduces the operational duration of the technology by eliminating the need for chemical disposal. This technology, by its very nature, enables the reduction of antibiotic release into the environment, and furthermore, it does not generate toxic substances, as is the case during the chlorination or ozonation processes. Furthermore, it enables operation at lower temperatures compared to the solar PF reaction. The development of this method is consistent with the objectives of sustainable development goals (UN, 2023). Cold plasma facilitates integration with green technologies, such as photovoltaics, which could result in a reduction in energy consumption (Li et al., 2021).

*4.4.5. Persulfate*

Persulfate (iron (III) sulfate) is a compound that is often used in wastewater treatment (Q. Zhang et al., 2015). Persulfate is a relatively inexpensive and effective coagulant that can be applied in wastewater treatment. However, PS concentration, the applied dose, and wastewater pH have to be controlled to guarantee the effectiveness of treatment. Persulfate can degrade up to 90.9% antibiotics from wastewater, depending on its concentration. Acidity and high temperature (up to 50°C) accelerate the degradation of sulfadiazine (Calcio Gaudino et al., 2021). The disinfection process should be monitored to prevent the formation of sediments that slow down the degradation of chemical compounds. Treated wastewater should be controlled to guarantee high degradation efficiency (Calcio Gaudino et al., 2021; Q. Zhang et al., 2015).

## 5. A comparison of the effectiveness of different methods for disinfecting hospital wastewater

The choice of the optimal wastewater disinfection method is a challenging task. The strengths and weaknesses of each method should be considered to select a technique that is most effective under specific conditions. Ozonation offers several advantages, including the absence of chemical by-products, rapid action, no greenhouse gas emissions, and a higher susceptibility of gram-negative bacteria to this treatment. Furthermore, ozonation induces oxidative stress, which enhances its antimicrobial efficacy. However, it also has disadvantages, such as high operational costs and the production of toxic by-products, including aldehydes (Anthony et al., 2019; Crini & Lichtfouse, 2019; Hilbig et al., 2020; Jäger et al., 2018). Another important method is chlorination, which is cost-effective, very successful at degrading ARGs. Furthermore, it generates no waste, consumes little energy, and is easy to control. However, it produces toxic by-products like chloroform and increases antibiotic resistance risks (Crini & Lichtfouse, 2019; Hilbig et al., 2020; J. Wang et al., 2020). The advantages and disadvantages of the other wastewater disinfection methods described in this article are presented in Table 8.

**Table 8.** Advantages and disadvantages of different wastewater disinfection methods.

| Method | Advantages | Disadvantages | Cost | References |
|--------|------------|---------------|------|------------|
| **UV** | - UV light inactivates ARGs by inhibiting RNA and DNA synthesis, which leads to cell death;<br>- No harmful or toxic by-products;<br>- UV light does not exert harmful effects on aquatic fauna;<br>- UV light does not contribute to antibiotic resistance.<br>- no chemicals<br>- simple automation | - Higher energy consumption;<br>- Increased greenhouse gas emissions;<br>- High cost;<br>- Disinfection equipment has to be regularly inspected, and it exerts a negative impact on the environment by generating large amounts of waste;<br>- UV light has the potential to induce a VBNC state in bacteria;<br>- UV treatment is not highly effective in turbid environments.<br>- no residual effect - risk of secondary contamination | from $0.01 to $0.05 per cubic meter | (Barbosa et al., 2021; Collivignarelli et al., 2021; Crini & Lichtfouse, 2019; Hilbig et al., 2020; Jäger et al., 2018) |
| **UV-LED** | - Increased production of ROS which promote the degradation of ARGs;<br>- No by-products | - High UV-LED doses can contribute to the excessive removal of intracellular genes that can slow down the degradation of extracellular ARGs (500 mL/cm$^3$);<br>- High initial installation cost. | from $0.005 to $0.03 per cubic meter | (Zhao et al., 2023). |
| **Chlorination** | - Low cost;<br>- High efficacy – chlorine penetrates intracellular spaces, damages the cell membrane and the cytoplasm in bacterial cells;<br>- Chlorination does not generate waste;<br>- Low energy consumption;<br>- High process stability;<br>- Easy control over the process | - Chlorination produces toxic substances such as chloroform;<br>- Chlorination increases the risk of antibiotic resistance;<br>- Chlorine is harmful for aquatic flora and fauna;<br>- Chlorine is an air pollutant;<br>- Chlorine-based disinfectants are difficult to store.<br>- ryzyko korozji instalacji<br>- nieskuteczność w kierunku *Cryptosporidium*, *Giardia* i form przetrwalnych | from $0.01 to $0.03 USD per cubic meter | (Crini & Lichtfouse, 2019; Hilbig et al., 2020; J. Wang et al., 2020) |
| **Ozonation** | - Rapid action;<br>- No greenhouse gas emissions;<br>- Gram-negative bacteria are more susceptible to ozonation;<br>- Ozonation induces oxidative stress;<br>- Rapid action. | - High cost;<br>- Ozonation produces toxic by-products such as aldehydes;<br>- Failure to fully mineralize products. | $0.0033 to $0.04 per cubic meter | (Anthony et al., 2019; Crini & Lichtfouse, 2019; Hilbig et al., 2020; Jäger et al., 2018) |
| **Gamma radiation** | - Does not require chemical compounds;<br>- Very high efficiency;<br>- Possibility of full automation. | - High energy consumption;<br>- Radiological hazards;<br>- Use mainly in the medical and pharmaceutical industries, less so in municipal treatment plants. | | (Crini & Lichtfouse, 2019) |
| **Oxidation with hydroxides** | - Does not require chemical compounds. | - High energy consumption;<br>- Greenhouse gas emissions;<br>- Generation of toxic by-products. | | (Crini & Lichtfouse, 2019) |
| **Photocatalysis** | - Additional chemicals are not required;<br>- Low risk of generation of by-products. | - Higher start-up and maintenance costs.<br>- Difficulty with catalyst recovery. | | (Crini & Lichtfouse, 2019) |
| **UV/PS** | - Low risk of generation of by-products and leaving chemical residues in treated water/wastewater;<br>- No health risks to the for personnel;<br>- Activity over a wide pH range compared to PF. | - High cost;<br>- Sulfate radicals have a short lifetime. | | (Crini & Lichtfouse, 2019) |
| **EC** | - Removes both organic and inorganic pollutants;<br>- Additional chemicals are not required;<br>- Low cost;<br>- Environmentally-friendly;<br>- pH does not need to be controlled;<br>- ARGs are more effectively removed when EC is combined with UV. | - EC requires electricity;<br>- EC generates sediments which have to be neutralized or managed;<br>- Additional treatment is required to remove iron ions, leading to costs increase. | | (L. Chen, Xu, et al., 2020; Crini & Lichtfouse, 2019) |
| **Nanoparticles** | - Highly effective in eliminating bacteria and fungi;<br>- Integration with membranes is possible;<br>- Operation over a wide range of pH and temperatures. | - Nanomaterials are potentially toxic for the environment and humans;<br>- The use of nanoparticles has to be strictly monitored;<br>- Lack of clear regulations and risk assessments. | From $0.05 to $0.1 per cubic meter | (Ezeuko et al., 2021a; Malakootian et al., 2019; Ren et al., 2018; T. Nguyen et al., 2021) |
| **Photo Fenton** | - Wide spectrum of light;<br>- The potential exists for the utilization of sunlight;<br>- The potential exists for integration with other technologies. | - Requires an acidic pH (2.8–3.5);<br>- Requires chemicals;<br>- Forms gel deposits. | | (Mosaka, 2023b; Sharma et al., 2023; Fenton, 2017; Foteins et al. 2018). |
| **Cold plasma** | - High efficiency ;<br>- No need for chemicals;<br>- Low temperature operation;<br>- Can be integrated with other methods. | - High costs;<br>- Surface operation;<br>- Use of high voltage current. | | (Gonçalves et al., 2025; Nguyen et al., 2021) |

## 6. The impact of hospital wastewater pre-treatment methods on degradation of antibiotics in hospital wastewater and their environmental influence

Wastewater disinfection is a crucial process that reduces antibiotic resistance and contributes to public health. Moreover, pathogens must be eliminated from wastewater to protect the environment and minimize the transmission of infectious diseases (The Regulation 1774/2002). Wastewater disinfection methods are similar to the techniques for degrading antimicrobials, and both goals can be achieved with the use of a single technological solution.

Various disinfection methods have been developed over the years, but chlorination and ozonation are most widely used to disinfect hospital wastewater due to the low cost of these treatment methods. However, chlorination and ozonation can lead to the production of toxic secondary products in wastewater. These substances directly affect the quality of hospital wastewater reaching WWTPs, and harmful compounds can be released to the environment with treated wastewater. In addition, disinfection and antibiotic degradation treatments are highly energy-intensive processes (UV), which increases the use of energy resources and contributes to greenhouse gas emissions (Crini & Lichtfouse, 2019; Mosaka et al., 2023a). Inadequate wastewater treatment and the excessive use of disinfectants can be counterproductive by increasing the concentrations of antibiotic-resistant microorganisms, which can pose a threat to the environment and public health (Bengtsson-Palme & Larsson, 2016). Hospital wastewater must be disinfected to eliminate pathogenic microorganisms, but process parameters, including the contact time and the dose of the applied disinfectant, have to be carefully selected to ensure that the treatment is effective. Inadequately treated wastewater creates a supportive environment for the emergence of new MDR bacteria that can also develop tolerance to disinfectants. These superbugs pose the greatest challenge for contemporary medicine, and their release into the environment is the direct cause of serious infections in humans. Effective wastewater treatment methods should not only eliminate ARB, but also degrade ARGs. Most disinfection techniques decrease microbial counts, but they are not equally effective in degrading ARGs, which could lead to the emergence of MDRB in hospital wastewater. It has been showed that as the disinfectant residual value rise, so does the acute toxicity. Furthermore, compared to residual chlorine (0.17 mg L$^{-1}$), residual acute toxicity of peracetic acid was greater (2.68 mg L$^{-1}$) (Collivignarelli et al., 2017).

The use of photocatalysts ($TiO_2$, ZnO), photosensitizers (porphyrins, methylene blue) (Planini et al., 2023) and electrode materials ($IrO_2$, BDD) (Okur et al. 2022) greatly increases disinfection effectiveness. There are various modifications of photocatalysts, however, the combination of $Fe_2O_3$-$TiO_2$ has no improvement in disinfection efficiency (García-Muñoz et al. 2025). However, there are methods which, when modified, provide effective disinfecting results, like $TiO_2$ ALPH (Castro-Rojas et al. 2025). However, variables such as pH (Amiri et al., 2010), temperature (Abu, 1994), organic content and inorganic ions (An et al. 2023) adversely affect the processes. For effective pathogen removal from wastewater, proper material selection and environmental management are essential.

In the context of enhancing the efficacy of existing disinfection processes, the utilization of advanced catalytic materials, such as plasmonic nanoparticles (Krueger et al., 2023, Kiriarachchi et al., 2018), which facilitate light absorption and electron transfer, holds considerable promise. This technique has the potential to enhance the efficiency of visible light utilization. Among the categories of new-generation catalysts are also Z-scheme photocatalysts, such as $BiVO_4$/g-$C_3O_4$ (Lin et al., 2023) and Ag/AgCl@ chiral $TiO_2$ (D. Wang et al., 2015), which allow for the generation of more efficient radicals at lower activation energies. It is also noteworthy to mention the utilization of "green" biocatalysts, which serve as a natural source of Fe (Qin et al., 2024) elements. These elements are crucial for numerous disinfection processes. Nevertheless, endeavors to amalgamate the aforementioned methodologies, which have been demonstrated to enhance removal efficiency, such as photo-electro-Fenton (Echeverry-Gallego et al., 2023) or plasma-catalysis (Fu et al., 20223, Ouzar et al. 2025), ought not to be disregarded.

When evaluating the environmental impact of disinfection, it is worth mentioning the Life Cycle Assessment (LCA) analysis tool, which is used to evaluate the environmental impact of disinfection and many other processes (Rashid et al., 2023). By using this method, it becomes easier to incorporate all relevant information on materials, energy, costs, benefits and as the impact on the human health and on the environment, into strategic planning and policy making. The most important aspects of LCA for hospital wastewater disinfection include the effect on public health and the condition of water, air and soil. LCA analyses should not ignore the energy and the resource consumption required for disinfection, also taking into account the type of technology used. An important aspect is the identification and assessment of the toxicity of disinfection by-products (Demir et al., 2024; Rashid et al., 2023). According to the LCA analysis which has been recently carried out, the comparison of different methods for removing ARB and ARGs shows that UV disinfection is a much more environmentally friendly and a more sustainable technology than $SO_2$ dechlorination and disinfection with chlorine gas/hypochlorite solutions. This aspect speaks in favour of ultraviolet radiation technology (Demir et al., 2024).

It should be noted that the effectiveness of the described pre-treatment methods is influenced by many factors, including the type and concentration of pollutants, type of microorganisms, and process parameters. Hospital wastewater is characterized by high concentrations of ARB, ARGs, and antibiotics, and appropriate disinfection methods have to be applied to ensure that these pollutants are eliminated and/or degraded to prevent or decrease the spread of AR in the environment.

## 7. The use of pre-treatment methods for degradation of antibiotics and ARB and ARGs elimination from hospital wastewater – summary and conclusions

Disinfection of hospital wastewater plays a key role in controlling the spread of AR. Hospital wastewater should be processed with the use of advanced disinfection techniques to eliminate or significantly reduce the transmission of AR in the natural environment. In this article, various disinfection methods were evaluated for their ability to eliminate ARB and degrade ARGs and antibiotics in hospital wastewater. The literature review was based on research studies focusing on disinfection methods that do not fully remove pollutants that pose a threat to human health and the environment. Life Cycle Assessment is a commonly employed method for evaluating the environmental impact of various processes, such as wastewater treatment techniques. It facilitates the analysis of potential environmental effects across different stages, including production, usage, and disposal (Brentner et al., 2011). The results of the conducted analysis indicate that UV-C radiation degrades ARB, but is less effective in eliminating ARGs. It should also be noted that UV-C treatment is an environmentally friendly process. However, the analysis of the UV-C LCA revealed that its environmental impact is primarily associated with electricity consumption during the photoreactor's operational phase. This electricity use provokes negative environmental effects, which may arise from coal combustion and the management of radioactive waste generated by nuclear power plants for imported electricity (Notarnicola et al. 2023). Ozonation combined with other disinfection techniques effectively inactivates ARGs and ARB and degrades antibiotics. One of the greatest advantages of ozonation is that it does not generate harmful chemicals in the environment, but the LCA analysis indicates that the environmental impact is significantly influenced by the energy consumption of $O_3$ (Maniakova et al., 2023). In turn, chlorination is an effective disinfection method, but it creates toxic by-products. Advanced oxidation processes effectively decompose organic matter, bacteria, and antibiotics, but they have to be regularly monitored to prevent the formation of dangerous by-products. Furthermore, it is imperative to comprehend the interplay between disinfection technologies and materials in order to optimize sanitation practices. The judicious selection of materials is instrumental in enhancing the efficacy and sustainability of disinfection processes. Future advances in this field may focus on the development of smart materials with improved antimicrobial properties, with a view to further improving disinfection performance in a range of applications. The following conclusions can be formulated based on a review of recent literature:

- ARGs do not degrade as effectively as ARB during pre-treatment processes, which increase the risk of the release of ARGs to the environment and the emergence of new ARB among environmental bacteria;
- The most effective method to degrade ARGs is ozonation but cold plasma is really promising when it comes to degrade antibiotics;
- The Photo-Fenton process is an effective and environmentally-friendly method of wastewater disinfection;

- An analysis of the literature indicates that the following combinations of disinfection methods are most effective in processing wastewater that is highly abundant in antibiotics and their metabolites: $H_2O_2/O_3$, UV/NiO, $TiO_2$/UV/ $H_2O_2$, and $TiO_2$/UV/$O_3$;
- The effectiveness of ARGs elimination/degradation depends on the methods of treatment used and on the analysis resistance gene, e.g. *sul* genes are more resistant to chlorination than *tet* genes;
- UV radiation combined with ozonation is highly effective in eliminating ARGs and ARB;
- Chlorine is an effective disinfectant when applied at a concentration of 16 mg $L^{-1}$ for 30 min, but chlorination poses numerous risks for the environment;
- TiO2 ALPH (Castro-Rojas et al. 2025) is a method that allows the optimization of reactor operation, allowing increased removal of ARBs, ARGs;
- Ozone applied at 1-15 g for 10 min eliminates ARB in 100%, but is not equally effective in degrading ARGs;
- UV-C radiation applied at 25.8 mJ $cm^{-2}$ for 2 min, seems to be the most effective in degrading ARGs.
- All disinfection methods must be tested experimentally.
- The LCA analysis points towards the advantages of UV disinfection over chlorination.

In the above review, many aspects of disinfection have been analyzed and it is difficult to say unequivocally that there is one universal solution that can achieve the best results in disinfecting hospital wastewater. Numerous modifications of the methods allow to introduce a wide range of technologies that reduce the number of ARB and ARGs, but each of them requires individual experimental verification.

## 8. Acknowledgements

This work was supported by the National Science Center, Poland [Grant No. 2022/45/B/NZ7/00793]. This research was funded in whole or in part by For the purpose of Open Access, the author has applied a CC-BY-SA 4.0 public copyright licence to any Author Accepted Manuscript (AAM) version arising from this submission.

The authors would like to express their gratitude to Zuzana Korzeniewska for her invaluable assistance in proofreading and improving the language of this manuscript. Her meticulous attention to detail and insightful suggestions have significantly enhanced the clarity and readability of the text.

## 9. Supplementary data
Supplementary data for this article can be found at:

## 10. References

Abu, Z. H. (1994). *Journal of Environmental Science and Health . Part A : Environmental Science and Engineering and Toxicology : Toxic / Hazardous Substances and Environmental Engineering Effect of temperature on the kinetics of wastewater disinfection using ultraviolet rad*. February 2015, 585–603. https://doi.org/10.1080/10934529409376056

Akbari, M. Z., Xu, Y., Lu, Z., & Peng, L. (2021). Review of antibiotics treatment by advance oxidation processes. *Environmental Advances*, 5, 100111. https://doi.org/10.1016/j.envadv.2021.100111

Aleksić, S., Gotvajn, A. Ž., Premzl, K., Kolar, M., & Turk, S. Š. (2021). Ozonation of amoxicillin and ciprofloxacin in model hospital wastewater to increase biotreatability. *Antibiotics*, 10(11). https://doi.org/10.3390/antibiotics10111407

Alexander, J., Knopp, G., Dötsch, A., Wieland, A., & Schwartz, T. (2016). Ozone treatment of conditioned wastewater selects antibiotic resistance genes, opportunistic bacteria, and induce strong population shifts. *Science of the Total Environment*, 559, 103–112. https://doi.org/10.1016/j.scitotenv.2016.03.154

Amiri, F., Mesquita, M. M. F., & Andrews, S. A. (2010). Disinfection effectiveness of organic chloramines , investigating the effect of pH. *Water Research*, 44(3), 845–853. https://doi.org/10.1016/j.watres.2009.09.004

Anthony, E. T., Ojemaye, M. O., Okoh, O. O., & Okoh, A. I. (2019). A critical review on the occurrence of resistomes in the environment and their removal from wastewater using apposite treatment technologies: Limitations, successes and future improvement. *Environmental Pollution*, 263, 113791. https://doi.org/10.1016/j.envpol.2019.113791

Arslan-Alaton, I., Karatas, A., Pehlivan, Ö., Koba Ucun, O., & Ölmez-Hancı, T. (2021). Effect of UV-A-assisted iron-based and UV-C-driven oxidation processes on organic matter and antibiotic resistance removal in tertiary treated urban wastewater. *Catalysis Today*, *361*(February), 152–158. https://doi.org/10.1016/j.cattod.2020.02.037

Auerbach, E. A., Seyfried, E. E., & McMahon, K. D. (2007). Tetracycline resistance genes in activated sludge wastewater treatment plants. *Water Research*, *41*(5), 1143–1151. https://doi.org/10.1016/j.watres.2006.11.045

Azuma, T., & Hayashi, T. (2021). Disinfection of Antibiotic-resistant Bacteria in Sewage and Hospital Effluent by Ozonation. *Ozone: Science and Engineering*, *43*(5), 413–426. https://doi.org/10.1080/01919512.2021.1906095

Azuma, T., Katagiri, M., Sekizuka, T., Kuroda, M., & Watanabe, M. (2022). Inactivation of Bacteria and Residual Antimicrobials in Hospital Wastewater by Ozone Treatment. *Antibiotics*, *11*(7), 1–13. https://doi.org/10.3390/antibiotics11070862

Azuma, T., Otomo, K., Kunitou, M., Shimizu, M., Hosomaru, K., Mikata, S., Ishida, M., Hisamatsu, K., Yunoki, A., Mino, Y., & Hayashi, T. (2019). Environmental fate of pharmaceutical compounds and antimicrobial-resistant bacteria in hospital effluents, and contributions to pollutant loads in the surface waters in Japan. *Science of the Total Environment*, *657*, 476–484. https://doi.org/10.1016/j.scitotenv.2018.11.433

Azuma, T., Usui, M., Hasei, T., & Hayashi, T. (2024). On-Site Inactivation for Disinfection of Antibiotic-Resistant Bacteria in Hospital Effluent by UV and UV-LED. *Antibiotics*, *13*(711), 1–21.

Baghal Asghari, F., Dehghani, M. H., Dehghanzadeh, R., Farajzadeh, D., Shanehbandi, D., Mahvi, A. H., Yaghmaeian, K., & Rajabi, A. (2021). Performance evaluation of ozonation for removal of antibiotic-resistant Escherichia coli and Pseudomonas aeruginosa and genes from hospital wastewater. *Scientific Reports*, *11*(1), 1–10. https://doi.org/10.1038/s41598-021-04254-z

Balachandran, S., Charamba, L. V. C., Manoli, K., Karaolia, P., Caucci, S., & Fatta-Kassinos, D. (2021). Simultaneous inactivation of multidrug-resistant Escherichia coli and enterococci by peracetic acid in urban wastewater: Exposure-based kinetics and comparison with chlorine. *Water Research*, *202*(June), 117403. https://doi.org/10.1016/j.watres.2021.117403

Balarak, D., & Mostafapour, F. K. (2019). Photocatalytic degradation of amoxicillin using UV/Synthesized NiO from pharmaceutical wastewater. *Indonesian Journal of Chemistry*, *19*(1), 211–218. https://doi.org/10.22146/ijc.33837

Barbosa, V., Morais, M., Silva, A., Delerue-Matos, C., Figueiredo, S. A., & Domingues, V. F. (2021). Comparison of antibiotic resistance in the influent and effluent of two wastewater treatment plants. *AIMS Environmental Science*, *8*(2), 101–116. https://doi.org/10.3934/environsci.2021008

Bengtsson-Palme, J., & Larsson, D. G. J. (2016). Concentrations of antibiotics predicted to select for resistant bacteria: Proposed limits for environmental regulation. *Environment International*, *86*, 140–149. https://doi.org/10.1016/j.envint.2015.10.015

Beretsou, V. G., Michael-Kordatou, I., Michael, C., Santoro, D., El-Halwagy, M., Jäger, T., Besselink, H., Schwartz, T., & Fatta-Kassinos, D. (2020). A chemical, microbiological and (eco)toxicological scheme to understand the efficiency of UV-C/H2O2 oxidation on antibiotic-related microcontaminants in treated urban wastewater. *Science of the Total Environment*, *744*, 140835. https://doi.org/10.1016/j.scitotenv.2020.140835

Bitter, K., Vlassakidis, A., Niepel, M., Hoedke, D., Schulze, J., Neumann, K., Moter, A., & Noetzel, J. (2017). Effects of Diode Laser , Gaseous Ozone , and Medical Dressings on Enterococcus faecalis Biofilms in the Root Canal Ex Vivo. *Hindawi BioMed Research International*, *2738*, 1–9. https://doi.org/10.1155/2017/6321850

Bojar, B., Sheridan, J., Beattie, R., Cahak, C., Liedhegner, E., Munoz-Price, L. S., Hristova, K. R., & Skwor, T. (2021). Antibiotic resistance patterns of Escherichia coli isolates from the clinic through the wastewater pathway. *International Journal of Hygiene and Environmental Health*, *238*(October), 113863. https://doi.org/10.1016/j.ijheh.2021.113863

Boševski, I., Kalčikova, G., Cerkovnik, J., & Žgajnar Gotvajn, A. (2020). Ozone as a Pretreatment Method for Antibiotic Contaminated Wastewater and Sludge. *Ozone: Science and Engineering*, *42*(2), 128–135. https://doi.org/10.1080/01919512.2019.1624149

Boudjema, N., Kherat, M., & Mameri, N. (2024). *Improved bacterial elimination in wastewater through electrocoagulation : hydrogen generation , adsorption of colloidal bacteria- fl ocks , and electric fi eld bactericidal action*. *14*(8), 744–757. https://doi.org/10.2166/washdev.2024.126

Bracamontes-Ruelas, A. R., Ordaz-Díaz, L. A., Bailón-Salas, A. M., Ríos-Saucedo, J. C., Reyes-Vidal, Y., & Reynoso-Cuevas, L. (2022). Emerging Pollutants in Wastewater, Advanced Oxidation Processes as an Alternative Treatment and Perspectives. *Processes*, *10*(5), 1–23. https://doi.org/10.3390/pr10051041

Brentner, L. B., Eckelman, M. J., & Zimmerman, J. B. (2011). Combinatorial life cycle assessment to inform process design of industrial production of algal biodiesel. *Environmental Science and Technology*, *45*(16), 7060–7067. https://doi.org/10.1021/es2006995

Bridges, D. F., Lacombe, A., & Wu, V. C. H. (2020). Integrity of the Escherichia coli O157:H7 Cell Wall and Membranes After Chlorine Dioxide Treatment. *Frontiers in Microbiology*, *11*(May), 1–11. https://doi.org/10.3389/fmicb.2020.00888

Calcio Gaudino, E., Canova, E., Liu, P., Wu, Z., & Cravotto, G. (2021). Degradation of antibiotics in wastewater: New

advances in cavitational treatments. *Molecules*, *26*(3). https://doi.org/10.3390/molecules26030617

Campo, N., De Flora, C., Maffettone, R., Manoli, K., Sarathy, S., Santoro, D., Gonzalez-Olmos, R., & Auset, M. (2020). Inactivation kinetics of antibiotic resistant Escherichia coli in secondary wastewater effluents by peracetic and performic acids. *Water Research*, *169*, 115227. https://doi.org/10.1016/j.watres.2019.115227

Chavoshan, S., Khodadadi, M., & Nasseh, N. (2020). Photocatalytic degradation of penicillin G from simulated wastewater using the UV/ZnO process: Isotherm and kinetic study. *Journal of Environmental Health Science and Engineering*, *18*(1), 107–117. https://doi.org/10.1007/s40201-020-00442-7

Chen, Y. di, Duan, X., Zhou, X., Wang, R., Wang, S., Ren, N. qi, & Ho, S. H. (2021). Advanced oxidation processes for water disinfection: Features, mechanisms and prospects. *Chemical Engineering Journal*, *409*(August 2020), 128207. https://doi.org/10.1016/j.cej.2020.128207

Chen, L., Xu, Y., Dong, X., & Shen, C. (2020). Removal of Intracellular and Extracellular Antibiotic Resistance Genes in Municipal Wastewater Effluent by Electrocoagulation. *Environmental Engineering Science*, *37*(12), 783–789. https://doi.org/10.1089/ees.2020.0189

Chen, L., Zhou, Z., Shen, C., & Xu, Y. (2020). Inactivation of antibiotic-resistant bacteria and antibiotic resistance genes by electrochemical oxidation/electroFenton process. *Water Science and Technology*, *81*(10), 2221–2231. https://doi.org/10.2166/wst.2020.282

Chhetri, R. K., Sanchez, D. F., Lindholst, S., Hansen, A. V., Sanderbo, J., Løppenthien, B. K., Eilkær, T., Gade, H., Skaarup, J., Kragelund, C., & Andersen, H. R. (2022). Disinfection of hospital-derived antibiotic-resistant bacteria at source using peracetic acid. *Journal of Water Process Engineering*, *45*(December 2021). https://doi.org/10.1016/j.jwpe.2021.102507

Chu, L., Wang, J., He, S., Chen, C., Wojnárovits, L., & Takács, E. (2021). Treatment of pharmaceutical wastewater by ionizing radiation: Removal of antibiotics, antimicrobial resistance genes and antimicrobial activity. *Journal of Hazardous Materials*, *415*(February). https://doi.org/10.1016/j.jhazmat.2021.125724

Collivignarelli, M. C., Abb, A., Gozio, E., & Benigna, I. (2017). *Disinfection in Wastewater Treatment Plants : Evaluation of Effectiveness and Acute Toxicity Effects*. https://doi.org/10.3390/su9101704

Collivignarelli, M. C., Abbà, A., Miino, M. C., Caccamo, F. M., Torretta, V., Rada, E. C., & Sorlini, S. (2021). Disinfection of wastewater by uv-based treatment for reuse in a circular economy perspective. Where are we at? *International Journal of Environmental Research and Public Health*, *18*(1), 1–24. https://doi.org/10.3390/ijerph18010077

Condon, S., Raso, J., Virto, R., & Man, P. (2005). Membrane Damage and Microbial Inactivation by Chlorine in the Absence and Presence of a Chlorine-Demanding Substrate. *Applied and Environmental Microbiology*, *71*(9), 5022–5028. https://doi.org/10.1128/AEM.71.9.5022

Crini, G., & Lichtfouse, E. (2019). Advantages and disadvantages of techniques used for wastewater treatment. *Environmental Chemistry Letters*, *17*(1), 145–155. https://doi.org/10.1007/s10311-018-0785-9

Cuerda-Correa, E. M., Alexandre-Franco, M. F., & Fernández-González, C. (2020). Advanced oxidation processes for the removal of antibiotics from water. An overview. *Water (Switzerland)*, *12*(1). https://doi.org/10.3390/w12010102

Cullen, P. J., Valdramidis, V. P., Tiwari, B. K., Patil, S., Bourke, P., & O'Donnell, C. P. (2010). Ozone processing for food preservation: An overview on fruit juice treatments. *Ozone: Science and Engineering*, *32*(3), 166–179. https://doi.org/10.1080/01919511003785361

Czekalski, N., Imminger, S., Salhi, E., Veljkovic, M., Kleffel, K., Drissner, D., Hammes, F., Bürgmann, H., & Von Gunten, U. (2016). Inactivation of Antibiotic Resistant Bacteria and Resistance Genes by Ozone: From Laboratory Experiments to Full-Scale Wastewater Treatment. *Environmental Science and Technology*, *50*(21), 11862–11871. https://doi.org/10.1021/acs.est.6b02640

Delaire, C., Genuchten, C. M. Van, Amrose, S. E., & Gadgil, A. J. (2016). Bacteria attenuation by iron electrocoagulation governed by interactions between bacterial phosphate groups and Fe ( III ) precipitates. *Water Research*, *103*, 74–82. https://doi.org/10.1016/j.watres.2016.07.020

Delgado-Blas, J. F., Valenzuela Agüi, C., Marin Rodriguez, E., Serna, C., Montero, N., Saba, C. K. S., & Gonzalez-Zorn, B. (2022). Dissemination Routes of Carbapenem and Pan-Aminoglycoside Resistance Mechanisms in Hospital and Urban Wastewater Canalizations of Ghana. *MSystems*, *7*(1). https://doi.org/10.1128/msystems.01019-21

Demir, M. Z., Guven, H., Ersahin, M. E., Ozgun, H., Pasaoglu, E., & Koyuncu, I. (2024). *Comparative Life Cycle Assessment of Four Municipal Water Disinfection Methods*.

Domínguez Henao, L., Turolla, A., & Antonelli, M. (2018). Disinfection by-products formation and ecotoxicological effects of effluents treated with peracetic acid: A review. *Chemosphere*, *213*, 25–40. https://doi.org/10.1016/j.chemosphere.2018.09.005

Dutta, V., Singh, P., Shandilya, P., Sharma, S., Raizada, P., Saini, A. K., Gupta, V. K., Hosseini-Bandegharaei, A., Agarwal, S., & Rahmani-Sani, A. (2019). Review on advances in photocatalytic water disinfection utilizing graphene and graphene derivatives-based nanocomposites. *Journal of Environmental Chemical Engineering*, *7*(3), 103132. https://doi.org/10.1016/j.jece.2019.103132

Dymaczewski, Z., Jeż-Walkowiak, J., Michałkiewicz, M., & Nr, M. M. S. (2019). Znaczenie procesu dezynfekcji w zapewnieniu bezpieczeństwa mikrobiologicznego wody przeznaczonej do spożycia przez ludzi. *Ochrona Środowiska*, *41*(1).

ECDC. (2022). Antimicrobial consumption in the EU / EEA ( ESAC-Net ). *European Centre for Disease Prevention and Control*, *November*.

Echeverry-Gallego, R. A., Espinosa-Barrera, P. A., Delgado-Vargas, C. A., Vanegas, J., Clavijo-Buriticá, D. C., Martínez-Pachón, D., & Moncayo-Lasso, A. (2023). The application of the photo-electro-Fenton process in the treatment of wastewater reduces the abundance of genes associated with pathogenicity factors, antibiotic resistance, and metabolism: A metagenomic analysis. *Journal of Environmental Chemical Engineering*, *11*(3), 109937.

Emmanuel, E., Keck, G., Blanchard, J. M., Vermande, P., & Perrodin, Y. (2004). Toxicological effects of disinfections using sodium hypochlorite on aquatic organisms and its contribution to AOX formation in hospital wastewater. *Environment International*, *30*(7), 891–900. https://doi.org/10.1016/j.envint.2004.02.004

Eramo, A., Morales Medina, W. R., & Fahrenfeld, N. L. (2017). Peracetic acid disinfection kinetics for combined sewer overflows: Indicator organisms, antibiotic resistance genes, and microbial community. *Environmental Science: Water Research and Technology*, *3*(6), 1061–1072. https://doi.org/10.1039/c7ew00184c

Esfandyari, Y., Saeb, K., Tavana, A., Rahnavard, A., & Fahimi, F. G. (2019). Effective removal of cefazolin from hospital wastewater by the electrocoagulation process. *Water Science and Technology*, *80*(12), 2422–2429. https://doi.org/10.2166/wst.2020.003

Esther, C. R., Kerr, A., & Gilligan, P. H. (2015). Detection of Mycobacterium abscessus from Deep Pharyngeal Swabs in Cystic Fibrosis. *Infection Control and Hospital Epidemiology*, *36*(5), 618–619. https://doi.org/10.1017/ice.2015.40

Ezeuko, A. S., Ojemaye, M. O., Okoh, O. O., & Okoh, A. I. (2021a). Potentials of metallic nanoparticles for the removal of antibiotic resistant bacteria and antibiotic resistance genes from wastewater: A critical review. *Journal of Water Process Engineering*, *41*(April), 102041. https://doi.org/10.1016/j.jwpe.2021.102041

Ezeuko, A. S., Ojemaye, M. O., Okoh, O. O., & Okoh, A. I. (2021b). Technological advancement for eliminating antibiotic resistance genes from wastewater: A review of their mechanisms and progress. *Journal of Environmental Chemical Engineering*, *9*(5), 106183. https://doi.org/10.1016/j.jece.2021.106183

Fenton, K. (2017). Innovative Waste Water Strategies in the Landscape: The Application of Green Infrastructure Principles in Cape Cod, Massachusetts.

Ferro, G., Guarino, F., Castiglione, S., & Rizzo, L. (2016). Antibiotic resistance spread potential in urban wastewater effluents disinfected by UV/H2O2 process. *Science of the Total Environment*, *560–561*, 29–35. https://doi.org/10.1016/j.scitotenv.2016.04.047

Finley, R. L., Collignon, P., Larsson, D. G. J., Mcewen, S. A., Li, X., Gaze, W. H., Reid-smith, R., Timinouni, M., Graham, D. W., & Topp, E. (2013). The Scourge of Antibiotic Resistance : The Important Role of the Environment. *Clinical Infectious Diseases*, *57*, 704–710. https://doi.org/10.1093/cid/cit355

Foteinis, S., Monteagudo, J. M., Durán, A., & Chatzisymeon, E. (2018). Environmental sustainability of the solar photo-Fenton process for wastewater treatment and pharmaceuticals mineralization at semi-industrial scale. *Science of the Total Environment*, *612*, 605-612.

Fu, Y., Wu, G., Geng, J., Li, J., Li, S., & Ren, H. (2019). Kinetics and modeling of artificial sweeteners degradation in wastewater by the UV/persulfate process. *Water Research*, *150*, 12–20. https://doi.org/10.1016/j.watres.2018.11.051

Fu, J., Xu, Y., Arts, E. J., Bai, Z., Chen, Z., & Zheng, Y. (2022). Viral disinfection using nonthermal plasma: A critical review and perspectives on the plasma-catalysis system. *Chemosphere*, *309*, 136655.

Geng, C., Liang, Z., Cui, F., Zhao, Z., Yuan, C., Du, J., & Wang, C. (2020). Energy-saving photo-degradation of three fluoroquinolone antibiotics under VUV/UV irradiation: Kinetics, mechanism, and antibacterial activity reduction. *Chemical Engineering Journal*, *383*(September), 123145. https://doi.org/10.1016/j.cej.2019.123145

Ghauch, A., Baalbaki, A., Amasha, M., El Asmar, R., & Tantawi, O. (2017). Contribution of persulfate in UV-254 nm activated systems for complete degradation of chloramphenicol antibiotic in water. *Chemical Engineering Journal*, *317*(2017), 1012–1025. https://doi.org/10.1016/j.cej.2017.02.133

Ghernaout, D., & Elboughdiri, N. (2020). Removing Antibiotic-Resistant Bacteria (ARB) Carrying Genes (ARGs): Challenges and Future Trends. *OALib*, *07*(01), 1–16. https://doi.org/10.4236/oalib.1106003

Giannakis, S., Rtimi, S., & Pulgarin, C. (2017). Light-assisted advanced oxidation processes for the elimination of chemical and microbiological pollution of wastewaters in developed and developing countries. *Molecules*, *22*(7). https://doi.org/10.3390/molecules22071070

Giannakis, S., Voumard, M., Rtimi, S., & Pulgarin, C. (2018). Applied Catalysis B : Environmental Bacterial disinfection by the photo-Fenton process : Extracellular oxidation or intracellular photo-catalysis ? *Applied Catalysis B: Environmental*, *227*(November 2017), 285–295. https://doi.org/10.1016/j.apcatb.2018.01.044

Gomes, J., Matos, A., Gmurek, M., Quinta-Ferreira, R. M., & Martins, R. C. (2019). Ozone and photocatalytic processes for pathogens removal from water: A review. *Catalysts*, *9*(1), 1–23. https://doi.org/10.3390/catal9010046

Gonçalves, J., Diaz, I., Kržišnik, D., & Žigon, J. (2025). *Killing Two Crises with One Spark : Cold Plasma for Antimicrobial Resistance Mitigation and Wastewater Reuse*. 1–15.

Gonz, Y., Gloria, G., Moeller-ch, G. E., & Vidal, G. (2023). *UV Disinfection Systems for Wastewater Treatment : Emphasis on Reactivation of Microorganisms*.

Gorito, A. M., Ribeiro, A. R. L., Rodrigues, P., Pereira, M. F. R., Guimarães, L., Almeida, C. M. R., & Silva, A. M. T. (2022). Antibiotics removal from aquaculture effluents by ozonation: chemical and toxicity descriptors. *Water Research*, *218*(April). https://doi.org/10.1016/j.watres.2022.118497

Guo, M. T., Yuan, Q. Bin, & Yang, J. (2013). Ultraviolet reduction of erythromycin and tetracycline resistant heterotrophic bacteria and their resistance genes in municipal wastewater. *Chemosphere*, *93*(11), 2864–2868. https://doi.org/10.1016/j.chemosphere.2013.08.068

Harif, T., Khai, M., & Adin, A. (2012). Electrocoagulation versus chemical coagulation : Coagulation / flocculation mechanisms and resulting floc characteristics. *Water Research*, *46*(10), 3177–3188. https://doi.org/10.1016/j.watres.2012.03.034

Hassaballah, A. H., Bhatt, T., Nyitrai, J., Dai, N., & Sassoubre, L. (2020). Inactivation of: E. coli, Enterococcus spp., somatic coliphage, and Cryptosporidium parvum in wastewater by peracetic acid (PAA), sodium hypochlorite, and combined PAA-ultraviolet disinfection. *Environmental Science: Water Research and Technology*, *6*(1), 197–209. https://doi.org/10.1039/c9ew00837c

Hembach, N., Schmid, F., Alexander, J., Hiller, C., Rogall, E. T., & Schwartz, T. (2017). Occurrence of the mcr-1 colistin resistance gene and other clinically relevant antibiotic resistance genes in microbial populations at different municipal wastewater treatment plants in Germany. *Frontiers in Microbiology*, *8*(JUL), 1–11. https://doi.org/10.3389/fmicb.2017.01282

Heß, S., & Gallert, C. (2015). Sensitivity of antibiotic resistant and antibiotic susceptible Escherichia coli, Enterococcus and Staphylococcus strains against ozone. *Journal of Water and Health*, *13*(4), 1020–1028. https://doi.org/10.2166/wh.2015.291

Hilbig, J., Boysen, B., Wolfsdorf, P., & Rudolph, K. (2020). *Economic evaluation of different treatment options for water reuse in industrial parks using modular cost functions*. 419–430. https://doi.org/10.2166/wrd.2020.032

Holt, P., Barton, G., & Mitchell, C. (2006). *ELECTROCOAGULATION AS A WASTEWATER TREATMENT*. *1956*.

Hong, Y., Zeng, J., Wang, X., Drlica, K., & Zhao, X. (2019). Post-stress bacterial cell death mediated by reactive oxygen species. *PANS*, *116*(20), 10064–10071. https://doi.org/10.1073/pnas.1901730116

Hou, A. ming, Yang, D., Miao, J., Shi, D. yang, Yin, J., Yang, Z. wei, Shen, Z. qiang, Wang, H. ran, Qiu, Z. gang, Liu, W. li, Li, J. wen, & Jin, M. (2019). Chlorine injury enhances antibiotic resistance in Pseudomonas aeruginosa through over expression of drug efflux pumps. *Water Research*, *156*, 366–371. https://doi.org/10.1016/j.watres.2019.03.035

Hsiao, S. S., Hsu, C. Y., Ananthakrishnan, B., Hsu, M. H., Chien, Y. T., Wang, L. P., & Tung, H. H. (2023). Ozone micron bubble pretreatment for antibiotic resistance genes reduction in hospital wastewater treatment. *Sustainable Environment Research*, *33*(1). https://doi.org/10.1186/s42834-023-00203-9

Huang, J. J., Xi, J., Hu, H. Y., Li, Y., Lu, S. Q., Tang, F., & Pang, Y. C. (2016). UV light tolerance and reactivation potential of tetracycline-resistant bacteria from secondary effluents of a wastewater treatment plant. *Journal of Environmental Sciences (China)*, *41*, 146–153. https://doi.org/10.1016/j.jes.2015.04.034

Huang, J. J., Xi, J. Y., Hu, H. Y., Tang, F., & Pang, Y. C. (2013). Inactivation and regrowth of antibiotic-resistant bacteria by PAA disinfection in the secondary effluent of a municipal wastewater treatment plant. *Biomedical and Environmental Sciences*, *26*(10), 865–868. https://doi.org/10.3967/bes2013.012

Huang, Z., Qi, Z., & Liu, C. (2023). Evaluation of the disinfection effect and mechanism of - SO 4 • − and - HO • UV / persulfate salts. *Environmental Science and Pollution Research*, 52380–52389. https://doi.org/10.1007/s11356-023-26120-3

Hubeny, J., Ciesielski, S., Harnisz, M., & Korzeniewska, E. (2021). Impact of Hospital Wastewater on the Occurrence and Diversity of Beta-Lactamase Genes During Wastewater Treatment with an Emphasis on Carbapenemase Genes : A Metagenomic Approach. *Frontiers in Environmental Science*, *9*(October), 1–12. https://doi.org/10.3389/fenvs.2021.738158

Igere, B. E., Okoh, A. I., & Nwodo, U. U. (2020). Wastewater treatment plants and release: The vase of Odin for emerging bacterial contaminants, resistance and determinant of environmental wellness. *Emerging Contaminants*, *6*, 212–224. https://doi.org/10.1016/j.emcon.2020.05.003

Ikehata, K. (2019). Recent Research on Ozonation By-products in Water and Wastewater Treatment : Formation , Control , Mitigation , and Other Relevant Topics. *Springer Nature Singapore Pte*.

Jäger, T., Hembach, N., Elpers, C., Wieland, A., Alexander, J., Hiller, C., Krauter, G., & Schwartz, T. (2018). Reduction of Antibiotic Resistant Bacteria During Conventional and Advanced Wastewater Treatment, and the Disseminated Loads Released to the Environment. *Frontiers in Microbiology*, *9*(October), 1–16. https://doi.org/10.3389/fmicb.2018.02599

Jin, M., Liu, L., Wang, D. ning, Yang, D., Liu, W. li, Yin, J., Yang, Z. wei, Wang, H. ran, Qiu, Z. gang, Shen, Z. qiang, Shi, D. yang, Li, H. bei, Guo, J. hua, & Li, J. wen. (2020). Chlorine disinfection promotes the exchange of antibiotic resistance genes across bacterial genera by natural transformation. *ISME Journal*, *14*(7), 1847–1856. https://doi.org/10.1038/s41396-020-0656-9

Kaliakatsos, A., Gounaki, I., Dokianakis, S., Maragkaki, E., Stasinakis, A. S., Gyparakis, S., Katsarakis, N., Manios, T., Fountoulakis, M. S., & Venieri, D. (2023). Treatment of hospital wastewater: emphasis on ecotoxicity and antibiotic resistance genes. *Journal of Chemical Technology and Biotechnology*, *January*. https://doi.org/10.1002/jctb.7329

Kalli, M., Noutsopoulos, C., & Mamais, D. (2023). The Fate and Occurrence of Antibiotic-Resistant Bacteria and Antibiotic Resistance Genes during Advanced Wastewater Treatment and Disinfection: A Review. *Water (Switzerland)*, *15*(11). https://doi.org/10.3390/w15112084

Kiriarachchi, H. D., Awad, F. S., Hassan, A. A., Bobb, J. A., Lin, A., & El-Shall, M. S. (2018). Plasmonic chemically modified cotton nanocomposite fibers for efficient solar water desalination and wastewater treatment. *Nanoscale*, *10*(39), 18531-18539.

Kitis, M. (2004). Disinfection of wastewater with peracetic acid: A review. *Environment International*, *30*(1), 47–55. https://doi.org/10.1016/S0160-4120(03)00147-8

Kocer, K., Boutin, S., Dalpke, A. H., Heeg, K., Mutters, N. T., & Nurjadi, D. (2020). Comparative genomic analysis reveals a high prevalence of inter-species in vivo transfer of carbapenem-resistance plasmids in patients with haematological malignancies. *Clinical Microbiology and Infection*, *26*(6), 780.e1-780.e8. https://doi.org/10.1016/j.cmi.2019.10.014

Korzeniewska, E., & Harnisz, M. (2018). Relationship between modification of activated sludge wastewater treatment and changes in antibiotic resistance of bacteria. *Science of the Total Environment*, *639*, 304–315. https://doi.org/10.1016/j.scitotenv.2018.05.165

Kowalska, J. (2016). *Kwas nadoctowy Peracetic acid*. *3*(3), 125–142.

Krueger, D., Graves, A., Chu, Y., Pan, J., Lee, S. E., & Park, Y. (2023). Integrated plasmonic gold nanoparticle dimer array for sustainable solar water disinfection. *ACS Applied Nano Materials*, *6*(7), 5568-5577.

Leggett, M. J., Schwarz, J. S., Burke, P. A., Mcdonnell, G., Denyer, S. P., & Maillard, J. (2016). Mechanism of Sporicidal Activity for the Synergistic Combination of Peracetic Acid and Hydrogen Peroxide. *American Society for Microbiology*, *110*(16), 1035–1039. https://doi.org/10.1128/AEM.03010-15.Editor

Li, D., & Gu, A. Z. (2019). Antimicrobial resistance: A new threat from disinfection byproducts and disinfection of drinking water? *Current Opinion in Environmental Science and Health*, *7*, 83–91. https://doi.org/10.1016/j.coesh.2018.12.003

Li, G., Bielicki, J. A., Ahmed, A. S. M. N. U., Islam, M. S., Berezin, E. N., Gallacci, C. B., Guinsburg, R., Eduardo, C., Vieira, R. S., Silva, A. R., Teixeira, C., Turner, P., Nhan, L., Orrego, J., Pérez, P. M., Qi, L., Papaevangelou, V., Chellani, H., Obaro, S., … Sharland, M. (2020). *Towards understanding global patterns of antimicrobial use and resistance in neonatal sepsis : insights from the NeoAMR network*. 26–31. https://doi.org/10.1136/archdischild-2019-316816

Li, M., Luo, K., & Xiong, Z. (2021). Design of adjustable high voltage pulse power supply driven by photovoltaic cells for cold plasma generation. In *2021 IEEE 4th International Electrical and Energy Conference (CIEEC)* (pp. 1-6). IEEE.

Lim, S., Shi, J. L., von Gunten, U., & McCurry, D. L. (2022). Ozonation of organic compounds in water and wastewater: A critical review. *Water Research*, *213*(January), 118053. https://doi.org/10.1016/j.watres.2022.118053

Lin, Z., Ye, S., Xu, Y., Lin, X., Qin, Z., Bao, J., & Peng, H. (2023). Construction of a novel efficient Z-scheme BiVO4/EAQ heterojunction for the photocatalytic inactivation of antibiotic-resistant pathogens: Performance and mechanism. *Chemical Engineering Journal*, *453*, 139747.

Liu, S. S., Qu, H. M., Yang, D., Hu, H., Liu, W. L., Qiu, Z. G., Hou, A. M., Guo, J., Li, J. W., Shen, Z. Q., & Jin, M. (2018). Chlorine disinfection increases both intracellular and extracellular antibiotic resistance genes in a full-scale wastewater treatment plant. *Water Research*, *136*, 131–136. https://doi.org/10.1016/j.watres.2018.02.036

Lou, W., Kane, A., Wolbert, D., Rtimi, S., & Assadi, A. A. (2017). Study of a photocatalytic process for removal of antibiotics from wastewater in a falling film photoreactor: Scavenger study and process intensification feasibility. *Chemical Engineering and Processing: Process Intensification*, *122*, 213–221. https://doi.org/10.1016/j.cep.2017.10.010

Luo, Y., Feng, L., Liu, Y., & Zhang, L. (2020). Disinfection by-products formation and acute toxicity variation of hospital wastewater under different disinfection processes. *Separation and Purification Technology*, *238*, 116405. https://doi.org/10.1016/j.seppur.2019.116405

Luprano, M. L., De Sanctis, M., Del Moro, G., Di Iaconi, C., Lopez, A., & Levantesi, C. (2016). Antibiotic resistance genes fate and removal by a technological treatment solution for water reuse in agriculture. *Science of the Total Environment*, *571*, 809–818. https://doi.org/10.1016/j.scitotenv.2016.07.055

Lutterbeck, C. A., Colares, G. S., Osbel, N. D., Silva, F. P., Kist, L. T., & Machado, L. (2020). *Hospital laundry wastewaters : A review on treatment alternatives , life cycle assessment and prognosis scenarios*. *273*. https://doi.org/10.1016/j.jclepro.2020.122851

Luukkonen, T., Teeriniemi, J., Prokkola, H., Rämö, J., & Lassi, U. (2014). Chemical aspects of peracetic acid based wastewater disinfection. *Water SA*, *40*(1), 73–80. https://doi.org/10.4314/wsa.v40i1.9

Malakootian, M., Mahdizadeh, H., Dehdarirad, A., & Amiri Gharghani, M. (2019). Photocatalytic ozonation degradation of ciprofloxacin using ZnO nanoparticles immobilized on the surface of stones. *Journal of Dispersion Science and Technology*, *40*(6), 846–854. https://doi.org/10.1080/01932691.2018.1485580

Maniakova, G., Polo López, M. I., Oller, I., Malato, S., & Rizzo, L. (2023). Ozonation Vs sequential solar driven processes as simultaneous tertiary and quaternary treatments of urban wastewater: A life cycle assessment comparison. *Journal of Cleaner Production*, *413*(December 2022). https://doi.org/10.1016/j.jclepro.2023.137507

Mao, C., Feng, Y., Wang, X., & Ren, G. (2015). Review on research achievements of biogas from anaerobic digestion. *Renewable and Sustainable Energy Reviews*, *45*, 540–555. https://doi.org/10.1016/j.rser.2015.02.032

Masoudi, F., Kamranifar, M., Safari, F., & Naghizadeh, A. (2019). Mechanism, kinetics and thermodynamic of penicillin g antibiotic removal by silica nanoparticles from simulated hospital wastewater. *Desalination and Water Treatment*, *169*, 333–341. https://doi.org/10.5004/dwt.2019.24657

McKinney, C. W., & Pruden, A. (2012). Ultraviolet disinfection of antibiotic resistant bacteria and their antibiotic resistance genes in water and wastewater. *Environmental Science and Technology*, *46*(24), 13393–13400. https://doi.org/10.1021/es303652q

Mehrabankhahi, G. K., Leili, M., Shokooni, R., Rahmani, A., Azarian, G., Khorram, N. S., & Razavi, M. (2023). The Optimization of Electrocoagulation Process Efficiency in the Removal of Amoxicillin Antibiotic and COD from Aqueous Solutions and Hospital Wastewater under Optimal Conditions: A Case Study of Alimoradian Hospital, Nahavand. *Scientific Journal of Kurdistan University of Medical Sciences*, *28*(1), 135–155.

Michael-Kordatou, I., Karaolia, P., & Fatta-Kassinos, D. (2018). The role of operating parameters and oxidative damage mechanisms of advanced chemical oxidation processes in the combat against antibiotic-resistant bacteria and resistance genes present in urban wastewater. *Water Research*, *129*, 208–230. https://doi.org/10.1016/j.watres.2017.10.007

Mohsin, M. K., & Mohammed, A. A. (2021). Catalytic ozonation for removal of antibiotic oxy-tetracycline using zinc oxide nanoparticles. *Applied Water Science*, *11*(1), 1–9. https://doi.org/10.1007/s13201-020-01333-w

Moreira, N. F. F., Narciso-da-Rocha, C., Polo-López, M. I., Pastrana-Martínez, L. M., Faria, J. L., Manaia, C. M., Fernández-Ibáñez, P., Nunes, O. C., & Silva, A. M. T. (2018). Solar treatment (H2O2, TiO2-P25 and GO-TiO2 photocatalysis, photo-Fenton) of organic micropollutants, human pathogen indicators, antibiotic resistant bacteria and related genes in urban wastewater. *Water Research*, *135*, 195–206. https://doi.org/10.1016/j.watres.2018.01.064

Mosaka, T. B. M., Unuofin, J. O., Daramola, M. O., Tizaoui, C., & Iwarere, S. A. (2023a). Inactivation of antibiotic-resistant bacteria and antibiotic-resistance genes in wastewater streams: Current challenges and future perspectives. *Frontiers in Microbiology*, *13*(January). https://doi.org/10.3389/fmicb.2022.1100102

Mosaka, T. B. (2023b). *Inactivation of Critically Ranked Carbapenem Resistant Bacteria and Genes in a Batch Atmospheric Plasma Reactor* (Master's thesis, University of Pretoria (South Africa)).

Müller, H., Sib, E., Gajdiss, M., Klanke, U., Lenz-Plet, F., Barabasch, V., Albert, C., Schallenberg, A., Timm, C., Zacharias, N., Schmithausen, R. M., Engelhart, S., Exner, M., Parcina, M., Schreiber, C., & Bierbaum, G. (2018). Dissemination of multi-resistant Gram-negative bacteria into German wastewater and surface waters. *FEMS Microbiology Ecology*, *94*(5), 1–11. https://doi.org/10.1093/FEMSEC/FIY057

Muñoz-castellanos, L. N., Borrego-loya, A., Villalba-bejarano, C. V., Orduño-cruz, N., Villezcas-villegas, G. P., & Rodríguez-, M. J. (2021). Chlorine and its importance in the inactivation of bacteria , can it inactivate viruses ? El cloro y su importancia en la inactivación de bacterias ,. *Mexican Journal of Phytopathology*, *39*(4), 198–206.

Nguyen, P. T. T., Nguyen, H. T., Tran, U. N. P., & Manh Bui, H. (2021). Removal of Antibiotics from Real Hospital Wastewater by Cold Plasma Technique. *Journal of Chemistry*, *2021*. https://doi.org/10.1155/2021/9981738

Nguyen, T. Q., Tung, K. L., Lin, Y. L., Dong, C. D., Chen, C. W., & Wu, C. H. (2021). Modifying thin-film composite forward osmosis membranes using various SiO2 nanoparticles for aquaculture wastewater recovery. *Chemosphere*, *281*, 130796.

Oguma, K., Izaki, K., & Katayama, H. (2013). Effects of salinity on photoreactivation of Escherichia coli after UV disinfection. *Journal of Water and Health*, *11*(3), 457–464. https://doi.org/10.2166/wh.2013.009

Omar, A., Almomani, F., & Qiblawey, H. (2024). *Advances in Nitrogen-Rich Wastewater Treatment : A Comprehensive Review of Modern Technologies*. 1–37.

Organization, W. H. (2017). *Guidelines for the prevention and control of carbapenem-resistant Enterobacteriaceae, Acinetobacter baumannii and Pseudomonas aeruginosa in health care facilities*.

Ouzar, A., Goutomo, B. T., Nam, K., & Kim, I. K. (2025). Efficient removal of tetracycline antibiotic by nonthermal plasma-catalysis combination process. *Environmental Engineering Research*, *30*(2).

Park, K. Y., Choi, S. Y., Lee, S. H., Kweon, J. H., & Song, J. H. (2016). Comparison of formation of disinfection by-products by chlorination and ozonation of wastewater effluents and their toxicity to Daphnia magna. *Environmental

*Pollution*, *215*, 314–321. https://doi.org/10.1016/j.envpol.2016.04.001

Phattarapattamawong, S., Chareewan, N., & Polprasert, C. (2021). Comparative removal of two antibiotic resistant bacteria and genes by the simultaneous use of chlorine and UV irradiation (UV/chlorine): Influence of free radicals on gene degradation. *Science of the Total Environment*, *755*, 142696. https://doi.org/10.1016/j.scitotenv.2020.142696

Pirsaheb, M., Hossaini, H., & Janjani, H. (2020). Reclamation of hospital secondary treatment effluent by sulfate radicals based–advanced oxidation processes (SR-AOPs) for removal of antibiotics. *Microchemical Journal*, *153*, 104430. https://doi.org/10.1016/j.microc.2019.104430

Planini, M., Crepulja, A., & Lus, M. (2023). *Photodynamic inactivation of multidrug- resistant strains of Klebsiella pneumoniae and Pseudomonas aeruginosa in municipal wastewater by tetracationic porphyrin and violet-blue light : The impact of wastewater constituents*. 1–23. https://doi.org/10.1371/journal.pone.0290080

Pokharel, S., Raut, S., & Adhikari, B. (2019). *Tackling antimicrobial resistance in low-income and middle--income countries*. 4–6. https://doi.org/10.1136/bmjgh-2019-002104

Popa, L. I., Gheorghe, I., Barbu, I. C., Surleac, M., Paraschiv, S., Măruţescu, L., Popa, M., Pîrcălăbioru, G. G., Talapan, D., Niţă, M., Streinu-Cercel, A., Streinu-Cercel, A., Oţelea, D., & Chifiriuc, M. C. (2021). Multidrug Resistant Klebsiella pneumoniae ST101 Clone Survival Chain From Inpatients to Hospital Effluent After Chlorine Treatment. *Frontiers in Microbiology*, *11*(January). https://doi.org/10.3389/fmicb.2020.610296

Proia, L., Anzil, A., Borrego, C., Farrè, M., Llorca, M., Sanchis, J., Bogaerts, P., Luis, J., & Servais, P. (2018). Occurrence and persistence of carbapenemases genes in hospital and wastewater treatment plants and propagation in the receiving river. *Journal of Hazardous Materials*, *358*(April), 33–43. https://doi.org/10.1016/j.jhazmat.2018.06.058

Pulicharla, R., Proulx, F., Behmel, S., & Jean-b, S. (2020). Trends in Ozonation Disinfection By-Products— Occurrence, Analysis and Toxicity of Carboxylic Acids. *Water*, *12*(756), 1–22. https://doi.org/doi:10.3390/w12030756

Qin, X., He, Y., Liu, S., & Shi, B. (2024). Persistent free radicals in natural organic matter activated by iron particles enhanced disinfection byproduct formation. *Water Research*, *266*, 122387.

Rafraf, I. D., Lekunberri, I., Sànchez-Melsió, A., Aouni, M., Borrego, C. M., & Balcázar, J. L. (2016). Abundance of antibiotic resistance genes in five municipal wastewater treatment plants in the Monastir Governorate, Tunisia. *Environmental Pollution*, *219*, 353–358. https://doi.org/10.1016/j.envpol.2016.10.062

Rangel, K., Cabral, F. O., Lechuga, G. C., Carvalho, P. R. S., Villas-b, M. H. S., Midlej, V., & De-simone, S. G. (2022). Detrimental Effect of Ozone on Pathogenic Bacteria. *Microorganisms*, *10*(40), 1–17. https://doi.org/https://doi.org/10.3390/microorganisms10010040

Rashid, S. S., Harun, S. N., Hanafiah, M. M., Razman, K. K., Liu, Y., & Tholibon, D. A. (2023). *Life Cycle Assessment and Its Application in Wastewater Treatment : A Brief Overview*. 1–31.

Reina, A. C., Martínez-Piernas, A. B., Bertakis, Y., Brebou, C., Xekoukoulotakis, N. P., Agüera, A., & Sánchez Pérez, J. A. (2018). Photochemical degradation of the carbapenem antibiotics imipenem and meropenem in aqueous solutions under solar radiation. *Water Research*, *128*, 61–70. https://doi.org/10.1016/j.watres.2017.10.047

Rekhate, C. V., & Srivastava, J. K. (2020). Recent advances in ozone-based advanced oxidation processes for treatment of wastewater- A review. *Chemical Engineering Journal Advances*, *3*(June), 100031. https://doi.org/10.1016/j.ceja.2020.100031

Ren, S., Boo, C., Guo, N., Wang, S., Elimelech, M., & Wang, Y. (2018). Photocatalytic Reactive Ultrafiltration Membrane for Removal of Antibiotic Resistant Bacteria and Antibiotic Resistance Genes from Wastewater Effluent [Research-article]. *Environmental Science and Technology*, *52*(15), 8666–8673. https://doi.org/10.1021/acs.est.8b01888

Richardson, S. D., & Postigo, C. (2015). Formation of DBPs: State of the Science. *ACS Symposium Series*, *1190*, 189–214. https://doi.org/10.1021/bk-2015-1190.ch011

Rizzo, L., Fiorentino, A., & Anselmo, A. (2013). Advanced treatment of urban wastewater by UV radiation: Effect on antibiotics and antibiotic-resistant E. coli strains. *Chemosphere*, *92*(2), 171–176. https://doi.org/10.1016/j.chemosphere.2013.03.021

Rizzo, L., Gernjak, W., Krzeminski, P., Malato, S., McArdell, C. S., Perez, J. A. S., Schaar, H., & Fatta-Kassinos, D. (2020). Best available technologies and treatment trains to address current challenges in urban wastewater reuse for irrigation of crops in EU countries. *Science of the Total Environment*, *710*, 136312. https://doi.org/10.1016/j.scitotenv.2019.136312

Rodrigues-Silva, C., Porto, R. S., dos Santos, S. G., Schneider, J., & Rath, S. (2019). Fluoroquinolones in hospital wastewater: Analytical method, occurrence, treatment with ozone and residual antimicrobial activity evaluation. *Journal of the Brazilian Chemical Society*, *30*(7), 1447–1457. https://doi.org/10.21577/0103-5053.20190040

Rodriguez-Mozaz, S., Vaz-Moreira, I., Varela Della Giustina, S., Llorca, M., Barceló, D., Schubert, S., Berendonk, T. U., Michael-Kordatou, I., Fatta-Kassinos, D., Martinez, J. L., Elpers, C., Henriques, I., Jaeger, T., Schwartz, T., Paulshus, E., O'Sullivan, K., Pärnänen, K. M. M., Virta, M., Do, T. T., … Manaia, C. M. (2020). Antibiotic residues in final effluents of European wastewater treatment plants and their impact on the aquatic environment. *Environment International*, *140*(April), 105733. https://doi.org/10.1016/j.envint.2020.105733

Rolbiecki, D., Korzeniewska, E., Czatzkowska, M., & Harnisz, M. (2022). The Impact of Chlorine Disinfection of Hospital Wastewater on Clonal Similarity and ESBL-Production in Selected Bacteria of the Family Enterobacteriaceae. *International Journal of Environmental Research and Public Health*, *19*(21). https://doi.org/10.3390/ijerph192113868

Rolbiecki, D., Paukszto, Ł., Krawczyk, K., Korzeniewska, E., Sawicki, J., & Harnisz, M. (2023). Chlorine disinfection modifies the microbiome, resistome and mobilome of hospital wastewater – A nanopore long-read metagenomic approach. *Journal of Hazardous Materials*, *459*(August). https://doi.org/10.1016/j.jhazmat.2023.132298

Rozporządzenie (WE) 1774/2002. (2002). *ROZPORZĄDZENIE (WE) 1774/2002*. *248*(1083), 1–89.

Serna-Galvis, E. A., Berrio-Perlaza, K. E., & Torres-Palma, R. A. (2017). Electrochemical treatment of penicillin, cephalosporin, and fluoroquinolone antibiotics via active chlorine: evaluation of antimicrobial activity, toxicity, matrix, and their correlation with the degradation pathways. *Environmental Science and Pollution Research*, *24*(30), 23771–23782. https://doi.org/10.1007/s11356-017-9985-2

Serna-Galvis, E. A., Ferraro, F., Silva-Agredo, J., & Torres-Palma, R. A. (2017). Degradation of highly consumed fluoroquinolones, penicillins and cephalosporins in distilled water and simulated hospital wastewater by UV254 and UV254/persulfate processes. *Water Research*, *122*, 128–138. https://doi.org/10.1016/j.watres.2017.05.065

Serna-Galvis, E. A., Jojoa-Sierra, S. D., Berrio-Perlaza, K. E., Ferraro, F., & Torres-Palma, R. A. (2017). Structure-reactivity relationship in the degradation of three representative fluoroquinolone antibiotics in water by electrogenerated active chlorine. *Chemical Engineering Journal*, *315*, 552–561. https://doi.org/10.1016/j.cej.2017.01.062

Shafaei, S., Klamerth, N., Zhang, Y., McPhedran, K., Bolton, J. R., & Gamal El-Din, M. (2017). Impact of environmental conditions on bacterial photoreactivation in wastewater effluents. *Environmental Science: Processes and Impacts*, *19*(1), 31–37. https://doi.org/10.1039/c6em00501b

Skandalis, N., Maeusli, M., Papafotis, D., Miller, S., Lee, B., Theologidis, I., & Luna, B. (2021). Environmental spread of antibiotic resistance. *Antibiotics*, *10*(6), 1–14. https://doi.org/10.3390/antibiotics10060640

Sharma, M., Tyagi, V. V., Chopra, K., Kothari, R., Singh, H. M., & Pandey, A. K. (2023). Advancement in solar energy-based technologies for sustainable treatment of textile wastewater: Reuse, recovery and current perspectives. *Journal of Water Process Engineering*, *56*, 104241.

Slipko, K., Reif, D., Schaar, H., Saracevic, E., Klinger, A., Wallmann, L., Krampe, J., Woegerbauer, M., Hufnagl, P., & Kreuzinger, N. (2022). Advanced wastewater treatment with ozonation and granular activated carbon filtration: Inactivation of antibiotic resistance targets in a long-term pilot study. *Journal of Hazardous Materials*, *438*(June), 129396. https://doi.org/10.1016/j.jhazmat.2022.129396

Sousa, J. M., Macedo, G., Pedrosa, M., Becerra-Castro, C., Castro-Silva, S., Pereira, M. F. R., Silva, A. M. T., Nunes, O. C., & Manaia, C. M. (2017). Ozonation and UV254nm radiation for the removal of microorganisms and antibiotic resistance genes from urban wastewater. *Journal of Hazardous Materials*, *323*, 434–441. https://doi.org/10.1016/j.jhazmat.2016.03.096

Stange, C., Sidhu, J. P. S., Toze, S., & Tiehm, A. (2019). Comparative removal of antibiotic resistance genes during chlorination, ozonation, and UV treatment. *International Journal of Hygiene and Environmental Health*, *222*(3), 541–548. https://doi.org/10.1016/j.ijheh.2019.02.002

Tabernacka, A. (2014). Biologiczne oczyszczanie wód podziemnych z chlorowcopochodnych etenu. *Ochrona Srodowiska*, *36*(1), 9–13.

Umar, M. (2022). From Conventional Disinfection to Antibiotic Resistance Control—Status of the Use of Chlorine and UV Irradiation during Wastewater Treatment. *International Journal of Environmental Research and Public Health*, *19*(3). https://doi.org/10.3390/ijerph19031636

Umar, M., Roddick, F., & Fan, L. (2019). Moving from the traditional paradigm of pathogen inactivation to controlling antibiotic resistance in water - Role of ultraviolet irradiation. *Science of the Total Environment*, *662*, 923–939. https://doi.org/10.1016/j.scitotenv.2019.01.289

United Nations. (2023). Goal 6 Clean water and sanitation. SDG Knowledge Portal. https://www.gov.pl/web/sdg-portal-wiedzy/cel-6-czysta-woda-i-warunki-sanitarne.

Viola, K. S., Rodrigues, E. M., Carlos, I. Z., Ramos, S. G., Ks, V., Em, R., Sg, R., Jm, G., & Faria, G. (2018). Cytotoxicity of peracetic acid : evaluation of effects on metabolism , structure and cell death. *International Endodonic Journal*, *51*, 264–277. https://doi.org/10.1111/iej.12750

von Sonntag, C., & von Gunten, U. (2012). Chemistry of Ozone in Water and Wastewater Treatment. In *Chemistry of Ozone in Water and Wastewater Treatment: From Basic Prinicples to Applications*. https://iwaponline.com/ebooks/book-pdf/650791/wio9781780400839.pdf

Wajahat, R., Yasar, A., Khan, A. M., Tabinda, A. B., & Bhatti, S. G. (2019). Ozonation and photo-driven oxidation of ciprofloxacin in pharmaceutical wastewater: Degradation kinetics and energy requirements. *Polish Journal of Environmental Studies*, *28*(3), 1933–1938. https://doi.org/10.15244/pjoes/90597

Wallmann, L., Krampe, J., Lahnsteiner, J., Radu, E., van Rensburg, P., Slipko, K., Wögerbauer, M., & Kreuzinger, N. (2021). Fate and persistence of antibiotic-resistant bacteria and genes through a multi-barrier treatment facility for direct potable reuse. *Water Reuse*, *11*(3), 373–390. https://doi.org/10.2166/wrd.2021.097

Wang, C., Lin, C. Y., & Liao, G. Y. (2020). Degradation of antibiotic tetracycline by ultrafine-bubble ozonation process. *Journal of Water Process Engineering*, *37*(December 2019), 101463. https://doi.org/10.1016/j.jwpe.2020.101463

Wang, D., Li, Y., Li Puma, G., Wang, C., Wang, P., Zhang, W., & Wang, Q. (2015). Dye-sensitized photoelectrochemical cell on plasmonic Ag/AgCl @ chiral TiO2 nanofibers for treatment of urban wastewater effluents, with simultaneous production of hydrogen and electricity. *Applied Catalysis B: Environmental*, *168–169*, 25–32. https://doi.org/10.1016/j.apcatb.2014.11.012

Wang, D., Yamaki, S., Kawai, Y., & Yamazaki, K. (2020). LWT - Food Science and Technology Sanitizing e ffi cacy and antimicrobial mechanism of peracetic acid against histamine-producing bacterium , Morganella psychrotolerans. *LWT - Food Science and Technology*, *126*(December 2019), 109263. https://doi.org/10.1016/j.lwt.2020.109263

Wang, H., Shi, W., Ma, D., Shang, Y., Wang, Y., & Gao, B. (2020). Formation of DBPs during chlorination of antibiotics and control with permanganate/bisulfite pretreatment. *Chemical Engineering Journal*, *392*, 123701. https://doi.org/10.1016/j.cej.2019.123701

Wang, H., Wang, J., Li, S., Ding, G., Wang, K., Zhuang, T., Huang, X., & Wang, X. (2020). Synergistic effect of UV/chlorine in bacterial inactivation, resistance gene removal, and gene conjugative transfer blocking. *Water Research*, *185*, 116290. https://doi.org/10.1016/j.watres.2020.116290

Wang, J., Shen, J., Ye, D., Yan, X., Zhang, Y., Yang, W., Li, X., Wang, J., Zhang, L., & Pan, L. (2020). Disinfection technology of hospital wastes and wastewater: Suggestions for disinfection strategy during coronavirus Disease 2019 (COVID-19) pandemic in China. *Environmental Pollution*, *262*, 114665. https://doi.org/10.1016/j.envpol.2020.114665

Wang, L., Li, Y., Ben, W., Hu, J., Cui, Z., Qu, K., & Qiang, Z. (2019). In-situ sludge ozone-reduction process for effective removal of fluoroquinolone antibiotics in wastewater treatment plants. *Separation and Purification Technology*, *213*(December 2018), 419–425. https://doi.org/10.1016/j.seppur.2018.12.062

Xiao, R., Liu, K., Bai, L., Minakata, D., Seo, Y., Kaya Göktaş, R., Dionysiou, D. D., Tang, C. J., Wei, Z., & Spinney, R. (2019). Inactivation of pathogenic microorganisms by sulfate radical: Present and future. *Chemical Engineering Journal*, *371*(April), 222–232. https://doi.org/10.1016/j.cej.2019.03.296

Xu, L., Zhang, C., Xu, P., & Wang, X. C. (2018). Mechanisms of ultraviolet disinfection and chlorination of Escherichia coli: Culturability, membrane permeability, metabolism, and genetic damage. *Journal of Environmental Sciences (China)*, *65*, 356–366. https://doi.org/10.1016/j.jes.2017.07.006

Yang, C., Wang, X., Zhang, L., Dong, W., Yang, C., Shi, X., Fan, Y., Wang, Y., Lv, H., Wang, W., & Zhao, Y. (2020). Investigation of kinetics and mechanism for the degradation of antibiotic norfloxacin in wastewater by UV/H2O2. *Journal of the Taiwan Institute of Chemical Engineers*, *115*, 117–127. https://doi.org/10.1016/j.jtice.2020.09.036

Yang, L., Wen, Q., Chen, Z., Duan, R., & Yang, P. (2019). Impacts of advanced treatment processes on elimination of antibiotic resistance genes in a municipal wastewater treatment plant. *Frontiers of Environmental Science and Engineering*, *13*(3). https://doi.org/10.1007/s11783-019-1116-5

Yuan, Q. Bin, Guo, M. T., & Yang, J. (2015). Fate of antibiotic resistant bacteria and genes during wastewater chlorination: Implication for antibiotic resistance control. *PLoS ONE*, *10*(3), 1–11. https://doi.org/10.1371/journal.pone.0119403

Zhang, Q., Chen, J., Dai, C., Zhang, Y., & Zhou, X. (2015). Degradation of carbamazepine and toxicity evaluation using the UV/persulfate process in aqueous solution. *Journal of Chemical Technology and Biotechnology*, *90*(4), 701–708. https://doi.org/10.1002/jctb.4360

Zhang, T., Wang, T., Mejia-tickner, B., Kissel, J., Xie, X., & Huang, C. (2020). Inactivation of Bacteria by Peracetic Acid Combined with Ultraviolet Irradiation: Mechanism and Optimization. *Environmental Science and Technology*, *54*, 9652–9661. https://doi.org/10.1021/acs.est.0c02424

Zhang, Y., Zhuang, Y., Geng, J., Ren, H., Zhang, Y., Ding, L., & Xu, K. (2015). Inactivation of antibiotic resistance genes in municipal wastewater effluent by chlorination and sequential UV/chlorination disinfection. *Science of the Total Environment*, *512–513*, 125–132. https://doi.org/10.1016/j.scitotenv.2015.01.028

Zhao, M., Zhou, X., Li, Z., Xu, G., Li, S., & Feng, R. (2023). The dynamics and removal efficiency of antibiotic resistance genes by UV-LED treatment: An integrated research on single- or dual-wavelength irradiation. *Ecotoxicology and Environmental Safety*, *263*(June). https://doi.org/10.1016/j.ecoenv.2023.115212

Zheng, J., Su, C., Zhou, J., Xu, L., Qian, Y., & Chen, H. (2017). Effects and mechanisms of ultraviolet, chlorination, and ozone disinfection on antibiotic resistance genes in secondary effluents of municipal wastewater treatment plants. *Chemical Engineering Journal*, *317*, 309–316. https://doi.org/10.1016/j.cej.2017.02.076

Zhou, C. shuang, Wu, J. wen, Dong, L. li, Liu, B. feng, Xing, D. feng, Yang, S. shan, Wu, X. kun, Wang, Q., Fan, J. ning, Feng, L. ping, & Cao, G. li. (2020). Removal of antibiotic resistant bacteria and antibiotic resistance genes in wastewater effluent by UV-activated persulfate. *Journal of Hazardous Materials*, *388*(January), 122070. https://doi.org/10.1016/j.jhazmat.2020.122070

# ARB inactivation, ARGs and antibiotics degradation in hospital wastewater

**Kornelia Stefaniak[1], Monika Harnisz[1], Magdalena Męcik[1] and Ewa Korzeniewska[1*]**

[1] Department of Water Protection Engineering and Environmental Microbiology, Faculty of Geoengineering, University of Warmia and Mazury in Olsztyn, Prawocheńskiego 1, 10-720 Olsztyn, Poland
* Corresponding author: ewa.korzeniewska@uwm.edu.pl

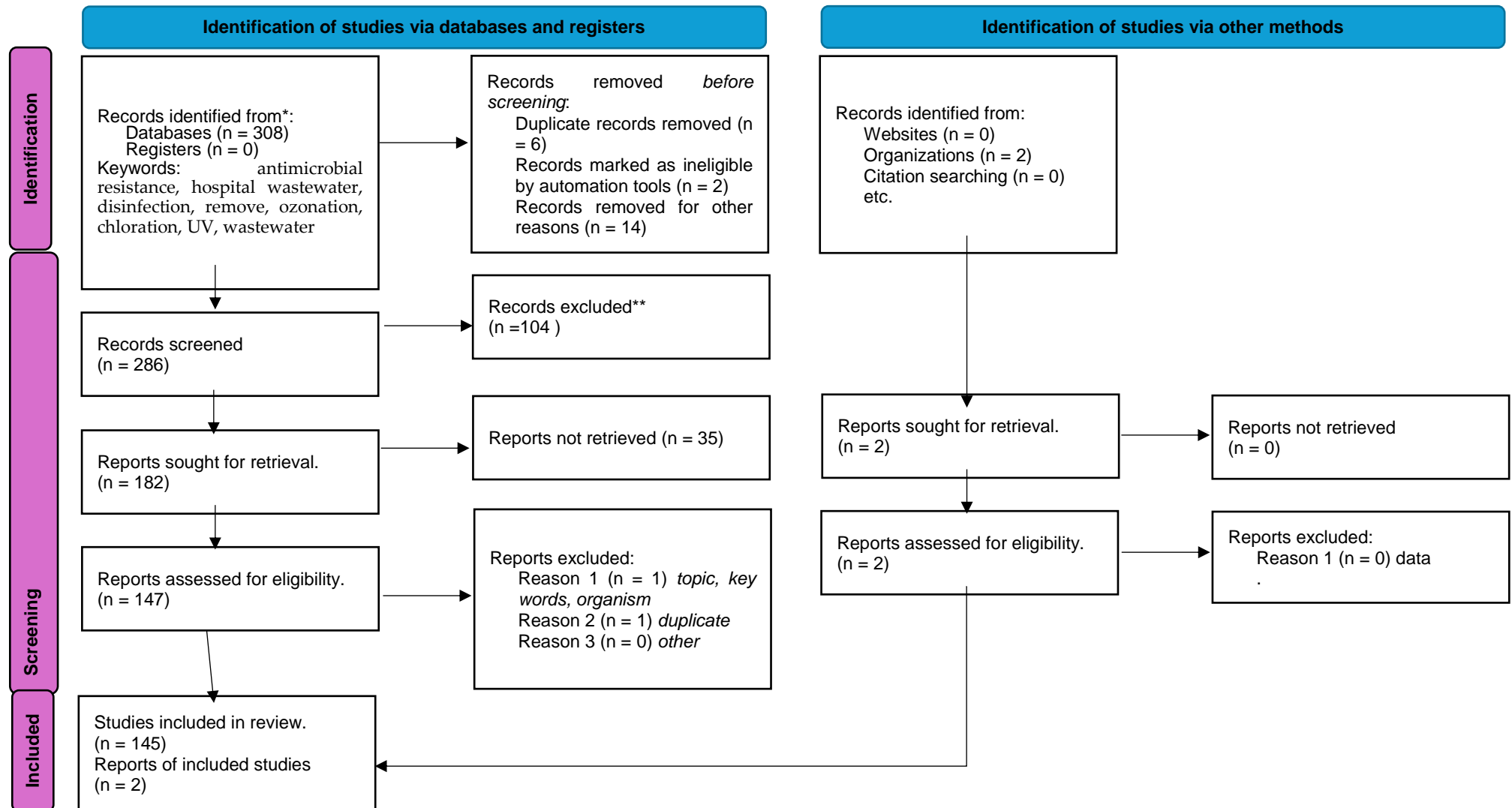

**Figure S1**. PRISMA flowchart showing the results of the publication's search and screening process for this review.

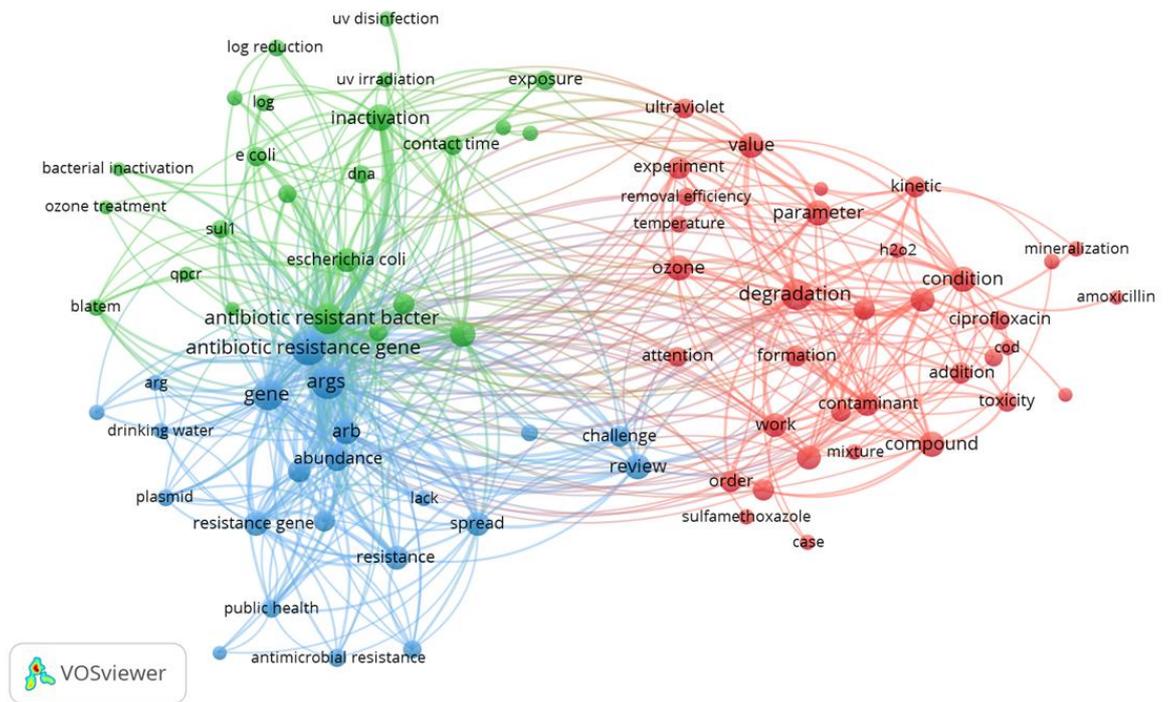

**Figure S2.** Network of keyword co-occurrence in articles containing the keyword "antibiotic resistance". The size of the nodes is proportional to the frequency of co-occurrence of a given keyword; node colors represent co-occurrence patterns in research articles published in each year of the analyzed period. The network was generated in VOSviewer (v. 1.6.19; 2023).

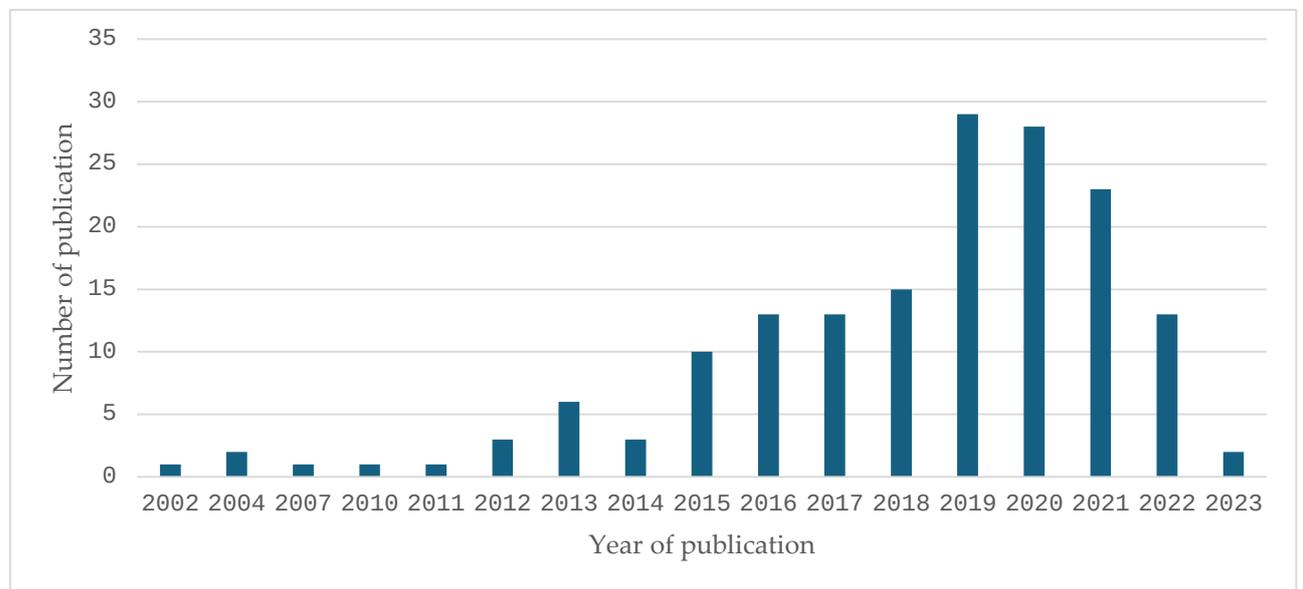

**Figure S3.** The number and date of publications used to develop this review.